\documentstyle[12pt,prl,aps,epsfig,citesort,psfig19,floats]{revtex}

\newcommand{\aleq}{\mbox{\ 
\raisebox{-.9ex}{$\stackrel{\textstyle<}{\sim}$}\ }}
\newcommand{\ageq}{\mbox{\
\raisebox{-.9ex}{$\stackrel{\textstyle >}{\sim}$}\ }}

\def\x{{\mbox{\boldmath$x$}}}
\def\u{{\mbox{\boldmath$u$}}}

\def\eps{{\epsilon}}

\def\begineq{\begin{equation}}
\def\endeq{\end{equation}}
\def\be{\begin{equation}}
\def\ee{\end{equation}}

\begin{document}
\bibliographystyle{prsty}
%\psdraft

\title{
Scaling in thermal convection: A unifying theory
}
\author{Siegfried Grossmann $^1$ and Detlef Lohse $^{1,2}$}
\address{
$^1$ Fachbereich Physik der Philipps-Universit\"at Marburg,\\
Renthof 6, D-35032 Marburg, Germany\\
$^2$ University of Twente,
Department of Applied Physics,\\
P.O. Box 217, 
7500 AE Enschede,
The Netherlands 
}

\date{\today}

\maketitle
\begin{abstract}
A systematic theory for the scaling of the Nusselt number $Nu$ and
of the Reynolds number $Re$ in strong Rayleigh-Benard convection is suggested
and shown to be compatible with recent experiments. It 
assumes a coherent large scale convection roll (``wind of turbulence'')
and 
is based on the dynamical
equations both in the bulk and in the boundary layers. Several regimes are
identified in the Rayleigh number $Ra$ versus Prandtl number $Pr$ phase space,
defined by whether the boundary layer or the bulk dominates the
global kinetic and thermal dissipation, respectively.
 The crossover between
the regimes is calculated. In the  regime which has
most frequently been studied in 
experiment ($Ra \aleq 10^{11}$)
the leading terms are 
$Nu\sim Ra^{1/4}Pr^{1/8}$, $Re \sim Ra^{1/2} Pr^{-3/4}$ for $Pr\aleq 1$
and 
$Nu\sim Ra^{1/4}Pr^{-1/12}$, $Re \sim Ra^{1/2} Pr^{-5/6}$ for $Pr\ageq 1$.
In most measurements
these laws 
are modified by additive corrections from the neighboring regimes so that
the impression of a slightly larger (effective) $Nu$ vs $Ra$ scaling exponent
can arise. 
The most important of the
 neighboring regimes towards large $Ra$ are 
a regime with scaling 
       $Nu \sim Ra^{1/2} Pr^{1/2} $,
       $Re\sim Ra^{1/2} Pr^{-1/2} $
for medium $Pr$ (``Kraichnan regime''),
a regime with scaling 
       $Nu \sim Ra^{1/5} Pr^{1/5} $,
       $Re\sim Ra^{2/5} Pr^{-3/5} $
for small $Pr$, 
a regime with $Nu \sim Ra^{1/3}$, 
              $ Re\sim Ra^{4/9} Pr^{-2/3}$ for larger $Pr$, 
and 
a regime with scaling 
       $Nu \sim Ra^{3/7} Pr^{-1/7} $,
       $Re\sim Ra^{4/7} Pr^{-6/7} $
for even larger $Pr$.
In particular, a linear combination 
of the 1/4 and the 1/3 power laws for $Nu$ with $Ra$,
$Nu = 0.27 Ra^{1/4} + 0.038 Ra^{1/3}$ 
(the prefactors follow from experiment),
mimicks a 2/7 power law exponent in a regime as large as ten decades. 
For very large $Ra$ 
the laminar shear  boundary layer is speculated to break down through
nonnormal-nonlinear transition to turbulence and another regime emerges. 
-- The presented theory is best summarized in the phase diagram figure
\ref{fig_sketch}.
\end{abstract}

\newpage

%----------------------------------------------------------------------

%\vspace{3cm}

\section{Introduction}
The early experiments on turbulent Rayleigh-Benard (RB) convection
in air cells with Prandtl number $Pr\approx 1$, summarized
by Davis \cite{dav22a}, showed a power law increase
of the Nusselt number $Nu$ with the Rayleigh number $Ra$,
namely, $Nu \sim Ra^\gamma$ with $\gamma = 1/4$.
{\footnote{
With the symbol ``$\sim$'' we mean ``scales as'' throughout
the text, {\it not} ``order of magnitude''.
The prefactors are determined in section IV. 
}}
However, in these
early experiments only relatively small Rayleigh numbers 
$Ra \aleq 10^8$ were achieved. Later, when RB 
experiments with larger Rayleigh numbers and in water cells
with $Pr\approx 7$ were done, the power law exponent $\gamma$
turned out to be larger than $1/4$. Malkus' elegant theory of
marginal stability, resulting in $\gamma = 1/3$, seemed to
describe those experiments \cite{mal54}.

In the late 80's, 
Libchaber et al.'s experiments
done at the University of Chicago on high Rayleigh number Rayleigh-Benard
convection in a helium gas cell with Prandtl number $Pr\approx 1$
revealed new and unexpected scaling for the
Nusselt number  as a function of the Rayleigh number, namely
$Nu\sim Ra^\gamma$ with $\gamma = 0.282\pm 0.006$ \cite{hes87,cas89}.
The Reynolds number $Re$,
characterizing the wind near the walls, i.e., the large eddy mean flow, 
 scaled as $Re\sim Ra^\alpha$ with $\alpha =
{0.491\pm 0.002}$
\cite{cas89}.
These results were reproduced and extended in many experiments and numerical
simulations
\cite{sol90,wu91a,pro91,wer93,chi93c,sig94,cio95,vil95,ker96,she96,tak96,cil96,cio97,xia97,cha97,qiu98,du98,lui98,ben99};
for review articles, which also summarize the results of earlier 
experimental, theoretical, and numerical work,
we refer to refs.\
\cite{sig94,cio97,zal98}. From all these experiments 
at first sight it seems that at least
the scaling
exponent $\gamma \approx 0.282\pm 0.006\approx 2/7$ 
is very robust.

Various theories were put forward to account for the scaling of the Nusselt
number $Nu$ and the Reynolds number $Re$
as functions of the Rayleigh number $Ra$ and the
Prandtl number $Pr$. These include
% two most prominent ones are
the Chicago mixing zone
model \cite{cas89} and the Shraiman-Siggia theory \cite{shr90},
 both reviewed in
refs.\ \cite{sig94,cio97,zal98}. The main result of the Chicago model is
$Nu\sim Ra^{2/7}$.
The Chicago group did not focus on the
$Pr$ dependence as only experiments with $Pr\approx 1$ were done \cite{cas89}.
Later, Cioni et al.\ \cite{cio97} added the $Pr$ dependence 
in the spirit of the
Chicago model and obtained
\begin{eqnarray}
Nu &\sim & Ra^{2/7} Pr^{2/7}, \label{nu_chicago}\\
Re_{fluct} &\sim & Ra^{3/7} Pr^{-4/7}.
\end{eqnarray}
Here, $Re_{fluct}$ refers to the velocity {\it fluctuations}
and not to the
large scale mean velocity (often denoted as the ``wind of turbulence'')
 as $Re$ does.
This Prandtl number dependence is only expected to hold for 
$Pr<1$ \cite{cio97,zal98}. 
 On the other hand, the Shraiman-Siggia
model, which assumes a turbulent boundary layer (BL) and a thermal
boundary layer nested therein (i.e., Shraiman and Siggia implicitly
assume a large enough Prandtl number), states
\begin{eqnarray}
Nu &\sim & Ra^{2/7} Pr^{-1/7}, \label{nu_ss}\\
\phantom{junkjunkjun}
Re &\sim & Ra^{3/7} Pr^{-5/7} \quad  \hbox{with logarithmic corrections}.
\end{eqnarray}
For an extension of the Shraiman-Siggia theory to position dependent shear
rates see ref.\ \cite{chi97}. 
The Prandtl number dependence of the Nusselt number resulting from 
 the Chicago model \cite{cas89} and from the Shraiman-Siggia
theory \cite{shr90} are not contradictory, 
as eq.\ (\ref{nu_chicago}) has been suggested for small
$Pr$ number fluids and eq.\ (\ref{nu_ss}) for the large $Pr$ number case.
Indeed, in the large $Pr$-limit Zaleski \cite{zal98} derives the 
same $Pr$-dependence 
eq.\ (\ref{nu_ss}) as in the Shraiman-Siggia theory
also from the Chicago model. 
However, both theories are based on rather different assumptions.

In recent years, 
the Prandtl number dependence of the Nusselt number
has been  measured and comparison with the two theories
\cite{cas89,shr90} became possible. 
The first experiments were done with water and helium RB cells. 
However, comparing $Nu$ in water and in helium convection  only
allows for a small variation of the Prandtl number.
From such experiments a small decrease of $Nu$ with increasing
$Pr$ at given $Ra$ was reported \cite{bel94}:
For $Ra=10^9$ Belmonte et al.\ \cite{bel94} measured
$Nu=76\pm 11$ for $Pr=0.7$ and 
$Nu=48\pm 6$ for $Pr=6.6$.  
Much larger $Pr$ variations are possible if RB convection in mercury
or liquid sodium is studied. 
Those experiments with a mercury
RB cell ($Pr=0.025$) by Rossby \cite{ros69},
by Takeshita et al.\ \cite{tak96}, and by Cioni et al.\
\cite{cio95,cio97} and with a liquid sodium RB cell ($Pr=0.005$)
by Horanyi et al.\ \cite{hor98} 
reveal that  $Nu$ {\it increases} with $Pr$, 
which is consistent with the Cioni et al.\ \cite{cio97} 
extension of the 
Chicago mixing zone model.

However, there seems to be  indication that
also the Chicago mixing zone model cannot account for all phenomena
observed in recent experiments: One of the most startling observations is that
there seems to be a small but significant  trend of the scaling exponent
$\gamma$ as a function of $Pr$. For $Pr \approx 5 - 7$ (water) one has
$\gamma = 0.28-0.293$ \cite{gar73,tan80,sig94,lui98} for $Ra$
up to $Ra\approx  10^9$
and an even larger
$\gamma \approx 1/3$ for larger $Ra\approx 10^9 - 10^{11}$ \cite{gol80};
for $Pr= 0.7 -1$ (helium gas) it is $\gamma = 0.282\pm 0.006$ \cite{cas89};
for $Pr=0.025$ (mercury) $\gamma = 0.247$ \cite{ros69} and
$\gamma = 0.26\pm 0.02$ \cite{cio97} (for $Ra< 10^9$) have been measured;
and for $Pr=0.005$
(liquid sodium) it is $\gamma = 0.25$ \cite{hor98}. 
Those and further 
experimental results are summarized in table \ref{tabneu}. 
All exponents between $0.25$ and $0.33$ have been measured! 
For thermal convection
in a water RB cell ($Pr\approx
 5-7$) with self similarly distributed balls on the
top and bottom wall, the scaling exponent can even be as large as
$\gamma = 0.45$, presumably depending on the ball size distribution
 \cite{cil99}.

\begin{table}[htb]
 \begin{center}
 \begin{tabular}{|c|c|c|c|c|}
 \hline
         Reference
       & fluid
       & $Pr$
       & $Ra$ range
       & $\gamma$
 \\
\hline
         Ashkenazi $\&$ Steinberg \cite{ash99}
       & SF$_6$
       & $1-93$
       & $10^9 - 10^{14}$
       & $0.30\pm 0.03$
\\
         Garon $\&$ Goldstein \cite{gar73}
       & H$_2$O
       & $5.5$
       & $10^7 - 3\cdot 10^9$
       & $0.293$
\\
         Tanaka $\&$ Miyata \cite{tan80}
       & H$_2$O
       & $6.8$
       & $3\cdot 10^7 - 4\cdot 10^9$
       & $0.290$
\\
         Goldstein $\&$ Tokuda \cite{gol80}
       & H$_2$O
       & $6.5$
       & $10^9 - 2\cdot 10^{11}$
       & $1/3$
\\
         Qiu $\&$ Xia  \cite{qiu98}
       & H$_2$O
       & $\approx 7$
       & $2\cdot 10^8 - 2\cdot 10^{10}$
       & $0.28$
\\
         Lui $\&$ Xia  \cite{lui98}
       & H$_2$O
       & $\approx 7$
       & $2\cdot 10^8 - 2\cdot 10^{10}$
       & $0.28 \pm 0.06$
\\
         Shen et al.\  \cite{she96}
       & H$_2$O
       & $\approx 7$
       & $8\cdot 10^7 - 7\cdot 10^{9}$
       & $0.281 \pm 0.015$
\\
         Threlfall \cite{thr75}
       & He
       & $0.8$
       & $4\cdot 10^5 - 2\cdot 10^{9}$
       & $0.280$
\\
         Castaing et al.\ \cite{cas89}
       & He
       & $0.7-1$
       & $\aleq 10^{11}$
       & $0.282\pm 0.006$
\\
         Wu $\&$ Libchaber \cite{wu91a}
       & He
       & $0.6-1.2$
       & $4\cdot 10^7 - 10^{12}$
       & $0.285$
\\
         Chavanne et al.\ \cite{cha97}
       & He
       & $0.6-0.73$
       & $3\cdot 10^7 - 10^{11}$
       & $2/7$
\\
         Davis \cite{dav22a}
       & air
       & $\approx 1$
       & $\aleq 10^8$
       & $0.25$
\\
         Rossby \cite{ros69}
       & Hg
       & $0.025$
       & $2\cdot 10^4 - 5 \cdot 10^5$
       & $0.247$
\\
         Takeshita et al.\ \cite{tak96}
       & Hg
       & $0.025$
       & $10^6 - 10^8$
       & $0.27$
\\
         Cioni et al.\ \cite{cio97}
       & Hg
       & $0.025$
       & $5\cdot 10^6 - 5\cdot 10^8$
       & $0.26\pm 0.02$
\\
         Cioni et al.\ \cite{cio97}
       & Hg
       & $0.025$
       & $4\cdot 10^8 - 2\cdot 10^9$
       & $0.20$
\\
         Glazier et al.\ \cite{gla99}
       & Hg
       & $0.025$
       & $2 \cdot 10^5 - 8 \cdot 10^{10} $
       & $0.29 \pm 0.01$
\\
         Horanyi et al.\ \cite{hor98}
       & Na
       & $0.005$
       & $\aleq 10^6$
       & $0.25$
\\
\hline
 \end{tabular}
 \end{center}
\caption[]{
Power law exponents $\gamma$ of the power law $Nu \sim Ra^\gamma$ 
for various experiments. The experiments were done with 
different aspect ratios, however, no strong dependence 
of the scaling exponent $\gamma$
on the aspect
ratio is expected (in contrast to the prefactors). 
}
\label{tabneu}
\end{table}

Next, a {\it breakdown} of the $\gamma\approx 2/7$ scaling regime at very large
$Ra$ has recently been observed, possibly towards a scaling regime $Nu \sim
Ra^{1/2}$, which has been predicted by Kraichnan decades ago \cite{kra62}.
For
$Pr=0.025$
Cioni et al.\ \cite{cio97} see the breakdown at $Ra \sim 2\cdot 10^9$
(and a startling small window with a local scaling exponent {\it
smaller} than $2/7$ for $Ra \sim 5\cdot 10^8 - 2\cdot 10^9$) while 
for $Pr=0.7-1.0$
Chavanne et
al.\ \cite{cha97} observe it at $Ra\sim 10^{11}$.
The transition around $Ra \approx 10^{11}$ in figure 3 of ref.\ \cite{sig94},
showing $Nu /Ra^{2/7}$ vs $Ra$ from Wu and Libchaber's data 
\cite{wu91b,wu91a}, may already be interpreted as the same breakdown.
On the other hand, Glazier et al.\ \cite{gla99}
do not observe such a transition. Thus, the experimental situation itself
is not yet clear.

All these observations and 
also the more intuitive rather than equation of motion based
approach
of ref.\ \cite{cas89} call for a re-examination
 and extension of the existing
scaling theories for thermal convection. Of course, a mathematically
rigorous derivation of $Nu(Ra,Pr)$ and $Re(Ra,Pr)$ is hardly possible.
The known
rigorous bounds 
overestimate the measured Nusselt numbers by more than one order of
magnitude and are only able to give the scaling exponent of the 
Kraichnan regime
$\gamma = 1/2$ \cite{how72,bus78,doe96}. 

Though in shortage of a strict mathematical derivation, the guideline of
the presented 
approach will be the dynamical equations for the velocity field
$\u (\x , t ) $,
the kinematic pressure
field $p(\x , t)$, 
 and the temperature field $\theta (\x , t)$,
\begin{eqnarray}
\partial_t u_i  + u_j \partial_j u_i &=& -\partial_i p
+\nu \partial_j^2 u_i +\beta g
\delta_{i3} \theta,
\label{eq5}
\\
\partial_t \theta  + u_j \partial_j
  \theta &=&  \kappa \partial_j^2 \theta , 
\label{eq6}
\end{eqnarray}
assisted by the appropriate boundary conditions at the bottom  wall $z=0$,
the top wall $z=L$,
 and the side walls of the cell. Here, $g$ is the gravitational
acceleration,
$\beta $ the 
isobaric 
thermal expansion coefficient, $\nu$ the kinematic viscosity,
$\kappa $ the thermal diffusivity, and $L$ the height of the RB cell; the
temperature difference between top and bottom walls is called $\Delta$.

The second feature of our approach 
introduced in section II
is that we try to be as systematic as
possible. 
We will be able to identify
four  
different main
scaling regimes for $Re$ and $Nu$ in the $Ra-Pr$ phase space,
depending on whether the BL or the bulk dominates the {\it global}
thermal and kinetic energy
dissipation, respectively. 
Three of the four regimes consist of two subregimes, depending on
whether the thermal BL or the viscous BL is thicker. 
We also calculate the validity range the scaling laws and 
make predictions on the stability of the
different regimes. 
In section III we compare the
power law exponents of the theory with experimental data.
In section IV we try to adopt the prefactors of the theory
to some experimental information and compare the resulting
prefactors to further experiments. 
Section V contains a summary and conclusions.

\section{Boundary layer vs bulk dominance of kinetic and thermal dissipation}
\subsection{Definitions}
The parameter space of RB convection is spanned by the Rayleigh and by the
Prandtl numbers,
\be
Ra = {\beta g L^3 \Delta \over \kappa \nu},\qquad
Pr={\nu \over \kappa}.
\ee
Our main focus is on the resulting Reynolds and Nusselt numbers,
\be
Re ={UL\over \nu }, \qquad Nu ={\left< u_z \theta\right>_A 
-\kappa \partial_3 \left< \theta \right>_A
 \over \kappa
 \Delta  L^{-1}},
 \ee
where
 $\left<.\right>_A$ denotes the average over (any) z-plane. Correspondingly,
 $\left<.\right>_V$ used below denotes the volume average.
$U$ is the mean large scale velocity
near the boundaries of the cell. It is the remainder of the convection
rolls 
which in the turbulent 
regime manifests itself as coherent large scale convection
flow, as first discovered by Krishnamurti and Howard \cite{kri81}
and later found by various groups
\cite{zoc90,wu91b,cas89,bel93,til93,bel94,sig94,xin96,qiu98}.
The existence of this ``wind of turbulence'' is one of the central
assumptions of our theory. 
We consider this to be a weak assumption, given the overwhelming
experimental evidence.  
The effect of the wind is twofold:
(i) In the range
between the wind and the cell wall a shear flow boundary layer will build up.
(ii) The wind  stirs the fluid in the bulk. 
In the presented theory we  consider 
the velocity
{\it fluctuations} in the bulk of the cell only as a {\it consequence} 
of the stirring by the large scale roll. Therefore, the Reynolds number
$Re$ based on the roll velocity rather than
the one based on the fluctuations $Re_{fluct}$ in the bulk is taken
as the more
appropriate one to theoretically describe the bulk turbulence.

\subsection{Decomposition of the energy dissipation}
The starting points of the present theory
are the kinetic and thermal dissipation rates
\begin{eqnarray}
\eps_u (\x , t ) &=& \nu (\partial_i u_j (\x , t))^2,\\
\eps_\theta (\x , t ) &=& \kappa (\partial_i \theta (\x , t))^2.
\end{eqnarray}
Their {\it global 
averages}
 $\left< \eps_u (\x ,t) \right>_V =\eps_u$ and 
 $\left< \eps_\theta (\x , t)\right>_V =\eps_\theta$
obey the following {\it rigorous} relations, which are easily
derivable from the equations of motion,  
see e.g.\ \cite{shr90,sig94}:
\begin{eqnarray}
\eps_u &=& {\nu^3\over L^4 } (Nu-1) Ra Pr^{-2},\label{eq11}\\
\eps_\theta &=& \kappa {\Delta^2 \over L^2} Nu. \label{eq12} 
\end{eqnarray}
Dissipation takes place both in the bulk of the flow and in the 
boundary layers. 
Near the walls thermal and kinetic boundary layers of thicknesses
$\lambda_\theta$ and $\lambda_u$ develop, which are determined by 
the thermal diffusivity $\kappa$ and the kinematic viscosity
$\nu$, respectively, and which are in general different, depending on $Pr$.
 They are defined on the basis of the
temperature and of the velocity profiles, respectively.
Whenever there exists a
thermal shortcut in the bulk 
due to the turbulent convective transport, 
the width of the thermal boundary layer is connected
with the Nusselt number by
\be
\lambda_\theta = {1\over 2}  L Nu^{-1}.
\ee
The thickness of the
kinetic boundary layer can be expressed in terms of the Reynolds number,
\be
\lambda_u \sim L Re^{-1/2}.
\label{eq14}
\ee
Here, we have assumed that there is  {\it laminar}
viscous flow of Blasius type (cf.\ sects.\ 39 and 41 of 
ref.\ \cite{ll87}) in the boundary layer;
the lateral extension $x$ of the BL
has been identified with the height $L$
of the cell,
reflecting that the wind organizes in the form of a large scale roll.
The transition to turbulence in  the boundary layers 
will be considered in subsection II-F. 
Though for large enough $Ra$ the total volume of the BLs is rather small, their
contribution to the global average
 dissipation may be considerable, as the velocity and
the temperature gradients in the BLs are much larger than in the bulk. 

In general, we decompose the globally averaged dissipation rates
into their BL and bulk contributions, 
\begin{eqnarray}
\eps_u &=& \eps_{u,BL} + \eps_{u,bulk}, \label{eq15}\\
\eps_\theta &=& \eps_{\theta ,BL} + \eps_{\theta , bulk},\label{eq16}
\end{eqnarray}
where
$\eps_{u,BL} = 
\int_{0\le z\le \lambda_u} +
\int_{L-\lambda_u \le z\le L} dz 
\nu (\partial_i u_j)^2/L = 
\nu  \left< (\partial_i u_j (\x \in BL ,t))^2\right>_{V}
$ 
is the viscous  dissipation taking place in the viscous BL,
$\eps_{\theta,BL} =
\int_{0\le z\le \lambda_\theta} +
\int_{L-\lambda_\theta \le z\le L} dz 
\kappa (\partial_i \theta )^2/L = 
\kappa  \left< (\partial_i \theta (\x \in BL ,t))^2\right>_{V}
$ is the thermal dissipation taking place in the thermal BL,
$\eps_{u,bulk} = 
\int_{\lambda_u \le z\le L- \lambda_u} 
dz 
\nu (\partial_i u_j)^2/L = 
\nu  \left< (\partial_i u_j (\x \in bulk ,t))^2\right>_{V}
$
is the viscous dissipation taking place in the bulk,
etc.

This kind of thinking immediately suggests the existence of four regimes:\\
(I) Both $\eps_u$ and $\eps_\theta$ are dominated by their BL contributions;\\
(II) $\eps_\theta$ is dominated by $\eps_{\theta,BL}$ and
   $\eps_u$ is dominated by $\eps_{u,bulk}$;\\
(III) $\eps_u$ is dominated by $\eps_{u,BL}$ and
   $\eps_\theta$ is dominated by $\eps_{\theta,bulk}$;\\
(IV) both $\eps_u$ and $\eps_\theta$ are bulk dominated.

For (relatively)
small $Ra$ the BLs are most extended, therefore regime I is 
expected. On the
other hand, for large $Ra$ the BLs are very thin and we
will expect regime IV, provided that the volume reduction of the BL is more
efficient than the dissipation increase in the BLs due to the growing
shear rate.
Next, for small $Pr$ the viscous BL is smaller than the thermal one,
$\lambda_u \ll \lambda_\theta$, and we expect regime II. Finally, for large $Pr$
it is $\lambda_u \gg \lambda_\theta$ and we have regime III. 

A priori it is not clear whether all four regimes can exist. However,
after input of some experimental information,
we will see 
that they are likely to  exist. 
We will also calculate the scaling 
of  the border lines
between the different regimes
in the $Ra-Pr$ phase space.
Of course, these lines do not indicate sharp transitions but the range
of the change of dominance.

\subsection{Estimate of bulk and BL contributions}
The next step is to estimate the various
contributions
$\eps_{u,BL}$,
$\eps_{\theta,BL}$,
$\eps_{u,bulk}$,
$\eps_{\theta,bulk}$
of the BL and the bulk dissipation 
from the dynamical equations
(\ref{eq5}) and (\ref{eq6}), expressing them as functions of $Nu$,
$Re$, $Ra$, and $Pr$. 

\noindent
{\underline{Bulk contributions:}}\\
We start with the {\it kinetic dissipation}. 
As already outlined above, the theory assumes that
 the bulk fluctuations with typical velocity $u_{fluct}$ 
{ originate} from the large scale coherent flow with velocity $U$. 
If there is no such 
``wind of turbulence'', the following estimates cease to be valid;
the Reynolds
number $Re=UL/\nu $ even cannot be defined properly.
Clearly, this will happen in the 
regime of very large Prandtl numbers where the flow is
suppressed by the strong 
viscosity. Therefore, there will be a transition
line $Pr(Ra)$ in phase space, 
defined by, say, $Re=50$, beyond which the theory no longer
holds; we will calculate this line in section II-E. 
Another limit of applicability is in the very small $Pr$ range.
Here, $\kappa$ is so large that the heat is molecularly conducted,
thus $Nu=1$. 
Other possible limitations of the basic assumptions of
 the theory will be discussed in section V. 

The assumption of the large scale velocity $U$ stirring
the bulk implies 
the picture of a turbulent energy cascade in the bulk which
in turn  suggests how to estimate
the {\it bulk} dissipation rates,
namely by balancing the dissipation with the large scale
convective term in the energy equations following from
(\ref{eq5}) and (\ref{eq6}),
\begin{equation}
\eps_{u,bulk} =
\nu  \left< (\partial_i u_j (\x \in bulk ,t))^2\right>_{V}
 \sim  {U^3\over L} = {\nu^3\over L^4} Re^3
\label{eq17}
\end{equation}
As argued before, we took 
as the relevant velocity scale  the wind  velocity $U$
and not the velocity fluctuations $u_{fluct}$ 
because it is  $U$ which {stirs} the fluid in the bulk.
This is a  key assumption of the theory, justified 
by intuition and by the results.
The presented theory does not  make any statement on the $Ra$-scaling of
the typical bulk fluctuations $u_{fluct}$. 

We explicitly remark that the findings on the $Re$ dependence of 
the energy dissipation rate in Taylor-Couette flow by Lathrop et al.
\cite{lat92} do {\it not} contradict eq.\ (\ref{eq17}).
Lathrop et al.\ \cite{lat92} found that $\eps_u L^4 \nu^{-3} Re^{-3}$
still depends on $Re$ even for large $Re$. However, their result
refers to the global $\eps_u$, {\it not} to $\eps_{u,bulk}$. -- 
Possible intermittency corrections are not taken into consideration
in eq.\ (\ref{eq17})  as they are at most small
\cite{gro95}.

We 
note that strictly speaking there should be a factor 
$(L-2\lambda_u )/L $ 
%$(L-2\lambda_\theta )/L $ 
on the rhs of eq.\ (\ref{eq17}), 
as the average
$\eps_{u,bulk}
= \left< \eps_u (\x \in bulk ,t)\right>_V$ 
%and 
%$\eps_{\theta,bulk} = \left< \eps_\theta (\x \in bulk ,t)\right>_V$
refers to the {\it whole} volume. However, it is assumed that the state is
already turbulent enough, i.e., $Ra$ large enough,  so that
$\lambda_u \ll L$. 
The validity of this assumption limits the scaling ranges to be derived. 

The estimate of the {\it thermal} bulk dissipation $\eps_{\theta , bulk}$ 
is slightly more complicated, as the velocity field $\u (\x , t )$ matters
in the dynamical eq.\ (\ref{eq6}) for the temperature. In particular,
it matters whether the kinetic BL, characterized by  a linear 
velocity profile, is nested in the thermal one or if it is 
the other way round.

For the former case ($\lambda_u < \lambda_\theta $, i.e., small $Pr$, 
%see
%figure \ref{sketch}a)
the thermal boundary layer can be estimated in complete analogy to 
eq.\ (\ref{eq17}) as 
\be
\eps_{\theta,bulk} =
\kappa  \left< (\partial_i \theta (\x \in bulk ,t))^2\right>_{V}
 \sim {U\Delta^2 \over L} = \kappa {\Delta^2 \over L^2}
Pr Re.\label{eq18}
\ee
Note that in correspondence with $U$ in   eq.\ (\ref{eq17}) 
 we took the large scale temperature difference
$\Delta$ in eq.\ (\ref{eq18}), not the typical temperature fluctuations
$\Delta_{fluct}$. Again, this is a key assumption, justified by the later
results.

For the latter case
($\lambda_u > \lambda_\theta $, i.e., large $Pr$, 
%see
%figure \ref{sketch}b) 
we must realize
that at the merging of the (linear)
 thermal BL into the thermal bulk the velocity
is 
{\it not} $U$ itself, but smaller by a factor
$\lambda_\theta / \lambda_u <1$.
 Therefore,
it is reasonable to assume that $U \lambda_\theta  /\lambda_u $ is 
the relevant velocity for the estimate of $\eps_{\theta , bulk}$, i.e.,
\be
\eps_{\theta,bulk} 
 \sim {\lambda_\theta \over \lambda_u } 
{U\Delta^2 \over L} = \kappa {\Delta^2 \over L^2}
Pr Re^{3/2} Nu^{-1}.
\label{eq18u}
\ee

%caption1
%\begin{figure}[htb]
%\setlength{\unitlength}{1.0cm}
%\begin{picture}(11,15)
%\put(1.0,7.5){\Large{(b)}}
%\put(1.0,15.){\Large{(a)}}
%\put(0.5,11)
%{\epsfig{figure=/home/lohse/turb/rb/paper50/sketch1.ps,width=4cm,angle=0}}
%\put(0.5,6.5)
%{\epsfig{figure=/usr/people/lohse/rb/paper50/sketch2.ps,width=4cm,angle=0}}
%\end{picture}
%\caption[]{
%Sketch of the boundary layers for low $Pr$ where $\lambda_u < \lambda_\theta$
%(a) and for large $Pr$ where  $\lambda_u > \lambda_\theta$ (b). 
%}
%\label{sketch}
%\end{figure}

\noindent
{\underline{BL contributions:}}\\
For $\eps_{u,BL}$ we follow an idea by Chavanne et al.\ 
\cite{cha97} and estimate, using eq.\ (\ref{eq14}), 
\be
\eps_{u,BL} = 
\nu  \left< (\partial_i u_j (\x \in BL ,t))^2\right>_{V}
\sim \nu {U^2 \over \lambda_u^2 }\cdot {\lambda_u\over L}
\sim {\nu^3 \over L^4 } Re^{5/2}.
\label{eq19}
\ee
Here, $U/\lambda_u$ characterizes the order of magnitude of $\partial_i u_j$
and 
the factor $\lambda_u /L$ accounts for the BL fraction of the total volume.
Again, this reasoning breaks down when there is no large scale
``wind of turbulence''. 
Correspondingly, we estimate
\be
\eps_{\theta,BL} = 
\kappa  \left< (\partial_i \theta (\x \in BL ,t))^2\right>_{V}
\sim \kappa {\Delta^2 \over \lambda_\theta^2 }\cdot {\lambda_\theta\over L}
\sim \kappa {\Delta^2 \over L^2 } Nu.
\label{eq20}
\ee
%The idea now is to only consider the dominating term on the rhs of eqs.\
%(\ref{eq15}) and (\ref{eq16}) 
%in the four corresponding regimes and to set it equal to the exact global
%averages
%$\eps_u$ (eq.\ \ref{eq11}) and  
%$\eps_\theta$ (eq.\ \ref{eq12}), respectively.
Eqs.\ (\ref{eq17}) to (\ref{eq20}) express the various dissipation
contributions (and thus the total dissipations $\epsilon_u$ 
and $\epsilon_\theta$,
(\ref{eq15}) and (\ref{eq16})) in terms of $Ra$, $Pr$, $Re$, and $Nu$.
If we insert 
eqs.\ (\ref{eq17}) to (\ref{eq20}) 
into the rigorous relations (\ref{eq11}) and (\ref{eq12}),
we obtain two equations, allowing to express $Nu$ and $Re$ in terms
of $Ra$ and $Pr$. If we only take the dominating contributions
$\epsilon_{BL}$ or 
$\epsilon_{bulk}$  in $\eps_u$ and $\eps_{\theta}$, respectively,
the formulae for the four regimes I, II, III, and IV are obtained,
describing pure scaling instead of superpositions. 

With this idea in mind, the scaling of the thermal boundary layer
dissipation (\ref{eq20}), though correct, does not give new information.
It coincides with the rigorous relation (\ref{eq12}). 
The physical reason is that the bulk is considered to provide a thermal
shortcut.
Therefore, we make use of 
the dynamics in the thermal BL in  more detail. We
approximate (systematically in order $1/Re$) eq.\ (\ref{eq6})
by the dominant terms
(cf.\ \cite{ll87} or \cite{shr90,cio97})
\be
u_x \partial_x \theta + u_z \partial_z \theta
=\kappa \partial_z^2 \theta \label{eq25}
\ee
in the thermal BL. 
Both terms on the lhs are of the same order of magnitude as can 
be concluded from the incompressibility condition
$\partial_x u_x +\partial_z u_z\approx 0$. 
In the lower subregimes with $\lambda_u < \lambda_\theta $
the velocity $u_x$ must be estimated by $U$, in the upper subregimes with
$\lambda_u > \lambda_\theta$ it is as argued above
$u_x \sim U \lambda_\theta /\lambda_u$.
%see figure \ref{sketch}. 
In addition, $\partial_x \sim 1/L$ and
$\kappa \partial_z^2 \sim \kappa /\lambda_\theta^2$.
 Therefore, for $\lambda_u < \lambda_\theta$ we finally get
\be
Nu \sim Re^{1/2} Pr^{1/2}
\label{eq26}
\ee
and for  $\lambda_u > \lambda_\theta $ we have 
\be
Nu \sim Re^{1/2} Pr^{1/3}.
\label{eq26u}
\ee
Eqs.\ (\ref{eq26}) and (\ref{eq26u}) replace the correct (but
useless) relation $\eps_\theta \sim
\eps_{\theta,BL}$ which does not add new information beyond (\ref{eq12}).

\subsection{Four regimes}

We will start with the $\eps_{\theta,bulk}$ dominated regimes
(III and IV).

\noindent
{\underline{Regime IV,
$\eps_u \sim \eps_{u,bulk}$ and 
$\eps_\theta \sim \eps_{\theta,bulk}$ (large $Ra$)}}\\
Depending on whether 
$\lambda_u$ is less or larger than $ \lambda_\theta$ 
we must use eq.\ (\ref{eq18}) 
or eq.\ (\ref{eq18u}), respectively, for the $\eps_{\theta , bulk}$-estimate. 
The former happens for low $Pr$, the latter for large $Pr$. Therefore,
we will give these two subregimes the index ``l'' for lower and ``u'' 
for upper. At what line $Pr(Ra)$ in phase space
$\lambda_u = \lambda_\theta$, 
the crossover from the $\lambda_u < \lambda_\theta$ to 
the $\lambda_u > \lambda_\theta $ will occur 
is not clear a priori.  We will later calculate this line
$\lambda_u = \lambda_\theta$ 
 with 
additional  experimental
information. 

In regime $IV_l$ we use (\ref{eq17}) for $\eps_u$ in (\ref{eq11}) and 
(\ref{eq18}) for $\eps_\theta$ in (\ref{eq12}) to obtain
\begin{eqnarray}
Nu &\sim & Ra^{1/2} Pr^{1/2}, \label{eq21}\\
Re &\sim & Ra^{1/2} Pr^{-1/2}. \label{eq22}
\end{eqnarray}
We recognize the asymptotic Kraichnan regime \cite{kra62}, just as
expected for large $Ra$ when both
thermal and kinetic energy dissipation are
bulk dominated. 
Note that other lines of arguments can also lead to eq.\ (\ref{eq21}),
see e.g.\ Kraichnan's work itself \cite{kra62}, Spiegel \cite{spi71},
or our reasoning in section II-F. Therefore, eq.\ (\ref{eq21})
seems to be quite robust. 
The physics of this regime is that the dimensional heat current
$Nu \kappa \Delta /L$ is independent of both $\kappa$ and $\nu$. 

In regime $IV_u$ we substitute as before (\ref{eq17}) for $\eps_u$ 
into eq.\ (\ref{eq11}) but now (\ref{eq18u}) instead of (\ref{eq18})
 for $\eps_\theta $ into eq.\  (\ref{eq12}) to obtain
\begin{eqnarray}
Nu &\sim & Ra^{1/3}, \label{eq21u}\\
Re &\sim & Ra^{4/9} Pr^{-2/3}. \label{eq22u}
\end{eqnarray}
The $Nu$ scaling is the one also following from the Malkus
 theory \cite{mal54}.

\noindent
{\underline{Regime III,
$\eps_u \sim \eps_{u,BL}$ and 
$\eps_\theta \sim \eps_{\theta,bulk}$ (large $Pr$)}}\\
Again we have to distinguish between the lower subregime $III_l$ 
with $\lambda_u < \lambda_\theta$ and the upper one $III_u$ 
with $\lambda_u > \lambda_\theta$. For $III_l$ we 
combine  (\ref{eq19}) with (\ref{eq11}) and 
(\ref{eq18}) with (\ref{eq12}) and get 
\begin{eqnarray}
Nu &\sim & Ra^{2/3} Pr^{1/3}, \label{eq23}\\
Re &\sim & Ra^{2/3} Pr^{-2/3}. \label{eq24}
\end{eqnarray}
This regime will turn out to be small and less important.
The more important one is $III_u$:
Combine (\ref{eq19}) with (\ref{eq11}) and 
(\ref{eq18u}) with (\ref{eq12}) to obtain
\begin{eqnarray}
Nu &\sim & Ra^{3/7} Pr^{-1/7}, \label{eq23u}\\
Re &\sim & Ra^{4/7} Pr^{-6/7}. \label{eq24u}
\end{eqnarray}
This regime may be observable for large enough $Pr$ when $\lambda_u
 \gg \lambda_\theta$.
To our knowledge up to date
this regime has neither been observed nor predicted.

Later, we will find hints for this regime $III_u$ 
in form of a subleading correction to describe the Chavanne et al.
data \cite{cha97}. It would be nice
to perform further experiments with large $Pr$ to be able to 
more cleanly identify
this postulated regime $III_u$.
We note that this regime is {\it not} in contradiction to 
Chan's \cite{cha71} upper
estimate $Nu \le const Ra^{1/3}$, holding 
in the infinite
$Pr$ limit (for fixed $Ra$), 
and also not in contradiction to Constantin and 
Doering's \cite{con99} {\it rigorous} upper bound
$Nu \le const Ra^{1/3}  (1+\log Ra)^{2/3}$,
holding in the same $Pr\to \infty $ limit.
The reason is that regime $III_u$ is for {\it finite}
$Pr$; if $Pr \to \infty$, also $Ra\to \infty$, if one wants
to stay in regime $III_u$.

\noindent
{\underline{Regime II,
$\eps_u \sim \eps_{u,bulk}$ and 
$\eps_\theta \sim \eps_{\theta,BL}$ (small $Pr$)}}\\
Regime $II_l$: 
Combining (\ref{eq17}) with (\ref{eq11}) gives together with 
(\ref{eq26})
\begin{eqnarray}
Nu &\sim & Ra^{1/5} Pr^{1/5}, \label{eq27}\\
Re &\sim & Ra^{2/5} Pr^{-3/5}. \label{eq28}
\end{eqnarray}
This regime should show up
for small enough
 $Pr$ when $\lambda_u \ll \lambda_\theta$. Indeed, Cioni et al.\
\cite{cio97} observed experimental hints for such a regime; also eqs.\
(\ref{eq27}) -- (\ref{eq28}) 
have already been derived by them in a similar way \cite{cio97}.
A 
power law 
$Nu \sim Ra^{1/5}$ 
was already suggested by 
Roberts \cite{rob79}. 

\noindent
Regime $II_u$: Because of the two competing conditions 
$\eps_\theta \sim \eps_{\theta , BL}$ (i.e., $Pr$ small) and
$\lambda_u > \lambda_\theta $ (i.e., $Pr$ large) such a subregime can 
at most be small. It will turn out later that it will probably 
not exist altogether. Nevertheless, for completeness we give the scaling
laws, resulting from taking now (\ref{eq26u}) and, as before,
inserting (\ref{eq17})
into (\ref{eq11}), namely
\begin{eqnarray}
Nu &\sim & Ra^{1/5}, \label{eq27u}\\
Re &\sim & Ra^{2/5} Pr^{-2/3}. \label{eq28u}
\end{eqnarray}

%caption1
\begin{figure}[htb]
\setlength{\unitlength}{1.0cm}
\begin{picture}(11,12)
%\put(0.5,12)
\put(0.5,0.5)
{\epsfig{figure=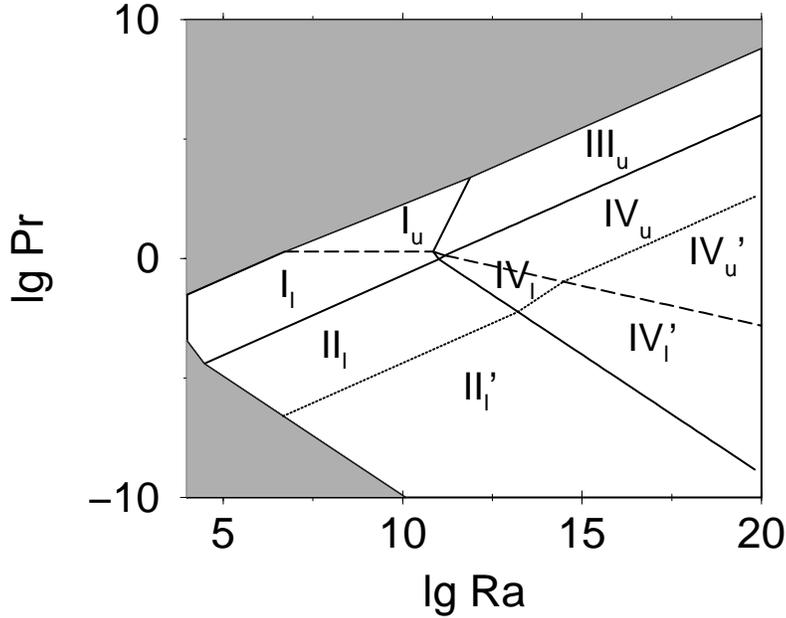,width=12cm,angle=-90}}
%{\epsfig{figure=/usr/people/lohse/rb/chavanne/phase_dia_new2.ps,width=12cm,angle=-90}}
\end{picture}
\caption[]{
Phase diagram in the $Ra-Pr$ plane. 
The power laws and the corresponding prefactors 
(to be determined in section IV) in the respective regimes
are summarized in 
 table \ref{tab_scal_laws}.
The tiny regime right of regime $I_l$ is regime $III_l$.
On the dashed line it is $\lambda_u = \lambda_\theta$. 
 In the shaded regime for large $Pr$ it is $Re\le 50$,
in the shaded regime for low $Pr$ we have $Nu = 1$. 
The dotted
line indicates the nonnormal-nonlinear onset of turbulence 
in the BL shear flow
discussed in subsection II-F.  The scaling in
regime $II_l^\prime$ is therefore as in the bulk 
dominated regime $IV_l$. --  
The power laws for the boundaries between the different regimes
are given in table \ref{tab_bounds}. 
}
\label{fig_sketch}
\end{figure}

 \begin{table}[htp]
 \begin{center}
 \begin{tabular}{|c|c|c|c|c|}
 \hline
          regime 
       &  dominance of
       &  BLs
       & $Nu$
       & $Re$
\\
\hline
         $I_l $ 
       & $\eps_{u,BL}$, $\eps_{\theta,BL}$
       & $\lambda_u < \lambda_\theta$
       & $0.27 Ra^{1/4} Pr^{1/8} $
       & $0.037 Ra^{1/2} Pr^{-3/4} $
\\         $I_u$ 
       & 
       & $\lambda_u > \lambda_\theta$
       & $0.33 Ra^{1/4} Pr^{-1/12} $
       & $0.039 Ra^{1/2} Pr^{-5/6} $
\\
\hline
         $II_l$
       & $\eps_{u,bulk}$, $\eps_{\theta,BL}$
       & $\lambda_u < \lambda_\theta$
       & $0.97 Ra^{1/5} Pr^{1/5} $
       & $0.47 Ra^{2/5} Pr^{-3/5} $
\\
         ($II_u$) 
       & 
       & $\lambda_u > \lambda_\theta$
       & ($\sim Ra^{1/5}  $)
       & ($ \sim Ra^{2/5} Pr^{-2/3} $)
\\
\hline
         $III_l$
       & $\eps_{u,BL}$, $\eps_{\theta,bulk}$
       & $\lambda_u < \lambda_\theta$
       & $6.43\cdot 10^{-6} Ra^{2/3} Pr^{1/3} $
       & $5.24\cdot 10^{-4} Ra^{2/3} Pr^{-2/3} $
\\
         $III_u$
       & 
       & $\lambda_u > \lambda_\theta$
       & $3.43\cdot 10^{-3} Ra^{3/7} Pr^{-1/7} $
       & $6.46\cdot 10^{-3} Ra^{4/7} Pr^{-6/7} $
\\
\hline
         $IV_l$
       & $\eps_{u,bulk}$, $\eps_{\theta,bulk}$
       & $\lambda_u < \lambda_\theta$
       & $4.43 \cdot 10^{-4} Ra^{1/2} Pr^{1/2} $
       & $0.036 Ra^{1/2} Pr^{-1/2} $
\\
         $IV_u$
       & 
       & $\lambda_u > \lambda_\theta$
       & $0.038 Ra^{1/3}  $
       & $0.16Ra^{4/9} Pr^{-2/3} $
\\
 \hline
 \end{tabular}
 \end{center}
\caption[]{
The power laws for $Nu$ and $Re$ of the presented theory,
including the prefactors which
are adopted to four pieces of experimental information in section IV. 
The exact
values of the prefactors depend also on how the Reynolds number is defined,
see the first paragraph of section IV.
Regime $II_u$ is put into brackets as it turns out that it does 
not exist for this choice of prefactors. 
}
\label{tab_scal_laws}
 \end{table}

\noindent
{\underline{Regime I,
$\eps_u \sim \eps_{u,BL}$ and 
$\eps_\theta \sim \eps_{\theta,BL}$}}\\
Regime $I_l$: 
This is the regime of (comparatively) small $Ra$ whose
scaling we obtain from using
(\ref{eq26}) and substituting
(\ref{eq19}) for $\eps_u$ in (\ref{eq11}), namely
\begin{eqnarray}
Nu &\sim & Ra^{1/4} Pr^{1/8}, \label{eq29}\\
Re &\sim & Ra^{1/2} Pr^{-3/4}. \label{eq30}
\end{eqnarray}
We argue that this is the regime whose scaling behavior has been observed
in almost all thermal turbulence experiments
\cite{hes87,cas89,sol90,wu91a,pro91,chi93c,sig94,cio95,cio97,tak96,cil96,xin96,xia97,cha97,qiu98,lui98},
but that 
in nearly all cases the pure scaling behavior (\ref{eq29}) and
(\ref{eq30}) has been polluted by sub-dominant contributions from the
neighboring regimes, as we will
elaborate in detail in the next section.

Remarkably, it is this power law $Nu\sim Ra^{1/4}$ which was the first
one suggested \cite{dav22a} and which has been well known in the
engineering literature for a long time \cite{fab95}. 
It also holds for two-dimensional convection in the low Prandtl number
limit \cite{cle81,bus81}.

\noindent
Regime
$I_u$: The scaling in $I_u$ is obtained from equation (\ref{eq26u}) and
combining (\ref{eq19}) with (\ref{eq11}), namely
\begin{eqnarray}
Nu &\sim & Ra^{1/4} Pr^{-1/12}, \label{eq29u}\\
Re &\sim & Ra^{1/2} Pr^{-5/6}. \label{eq30u}
\end{eqnarray}
Note that the $Ra$ dependence is the same as in $I_l$, but now  $Nu$ 
{\it decreases} with increasing $Pr$. 
This behavior is physically to be expected because due to increasing
$\nu$ the convective heat transport is more and more reduced. 
And indeed, 
such a crossover
from increase to decrease of $Nu$ with $Pr$ 
has been  observed
in experiment. 
It will later give us the opportunity to determine the
transition line $\lambda_u = \lambda_\theta$.

%caption1
\begin{figure}[htb]
\setlength{\unitlength}{1.0cm}
\begin{picture}(11,12)
%%\put(0.5,12)
\put(0.5,0.5)
{\psfig{figure=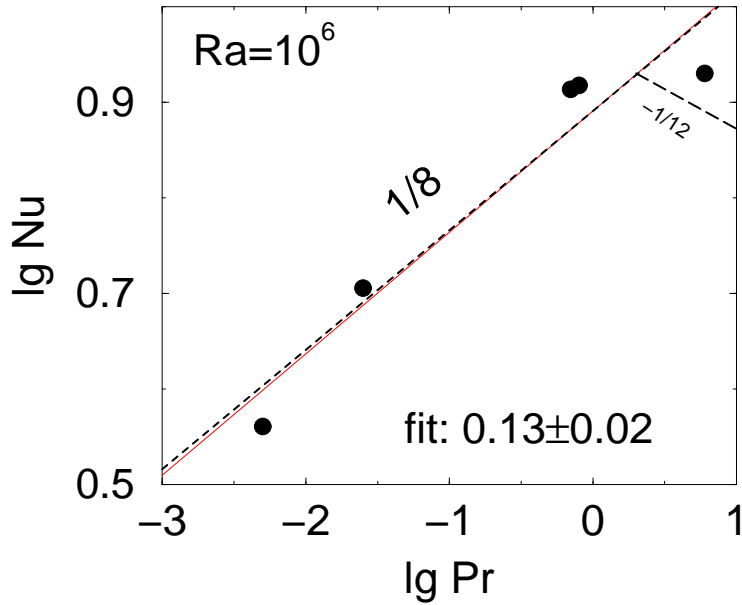,width=12cm,angle=-90}}
%{\psfig{figure=/usr/people/lohse/rb/cioni/nu_vs_pr_log.eps,width=12cm,angle=-90}}
%{\psfig{figure=/vhe/therm2/mnt/wagner/lohse/rb/cioni/nu_vs_pr_log.eps,width=12cm,angle=-90}}
%{\psfig{figure=[detlef.rb.slides]NU_VS_pr_log.PS,width=12cm,angle=-90}}
\end{picture}
\caption[]{
$Nu$ vs $Pr$ for fixed $Ra=10^6$.
The fit is based on all points. 
The expected power law $Nu \sim Pr^{1/8}$ in regime $I_l$ is also shown 
(dashed).
For larger $Pr \ge 2$ one enters 
regime $I_u$ where $Nu \sim Pr^{-1/12}$ (dashed) is expected. 
}
\label{fig_nu_vs_pr}
\end{figure}

The  scaling of this crossover line
can already be determined here from equating $\lambda_\theta
\sim L/Nu$ and $\lambda_u \sim L/\sqrt{Re}$ in the respective regimes.
We obtain
\begin{eqnarray}
Pr_{\lambda}^{I_l-I_u} &\sim & Ra^0 , \label{tll1}\\
Pr_{\lambda}^{II_l-II_u} &\sim & Ra^0 , \label{tll2}\\
Pr_{\lambda}^{III_l-III_u} &\sim & Ra^{-1/2}, \label{tll3}\\
Pr_{\lambda}^{IV_l-IV_u} &\sim & Ra^{-1/3}. \label{tll4}
\end{eqnarray}
Because the line $\lambda_u = \lambda_\theta$ obeys $Pr = const$ in 
regime $I$ it either is above or below the common corner point of all four
regimes. Therefore it can go either through
regime $III$ or through regime $II$, but not through both. 
Thus either regime $II_u$ will exist or regime $III_l$,
never both of them.

We now calculate
the scaling of the boundaries between the other 
different domains in the $Ra-Pr$ phase space. The boundary between I and II is
obtained by equating
$\eps_{u,BL} \sim \eps_{u,bulk}$, those between I and III by equating
$\eps_{\theta,BL} \sim \eps_{\theta,bulk}$, etc. 
The results are
\begin{eqnarray}
Pr_{trans}^{I_l-II_l} &\sim & Ra_{trans}^{2/3} , \label{eq31}\\
Pr_{trans}^{I_l-III_l} &\sim & Ra_{trans}^{-2} , \label{eq32}\\
Pr_{trans}^{III_l-IV_l} &\sim & Ra_{trans}^{1} , \label{eq33}\\
Pr_{trans}^{II_l-IV_l} &\sim & Ra_{trans}^{-1} , \label{eq34}\\
Pr_{trans}^{I_u-III_u} &\sim & Ra_{trans}^{3} , \label{eq32u}\\
Pr_{trans}^{III_u-IV_u} &\sim & Ra_{trans}^{2/3} . \label{eq33u}
\end{eqnarray}
Note that all these lines indicate the range of smooth crossover in the
dominance of either the BL or the bulk dissipation.

The phase diagram in $Ra-Pr$ phase space with the various regimes
and crossovers is shown in figure
\ref{fig_sketch}, anticipating the prefactors of the power
laws, whose exponents we have evaluated up to now. 
We will determine the prefactors in
eqs.\ (\ref{eq26}) -- (\ref{eq33u}) in section IV 
from  four pieces of experimental information. 
These experimental informations all come from experiments with an aspect
ratio of the RB cell of the order of 1. We expect the prefactors to
depend on the aspect ratio; therefore, all prefactors given in this paper
only refer to  aspect ratio order of 1 experiments.

\subsection{Range of validity of power laws}
What is the range of validity of the power laws summarized in table
\ref{tab_scal_laws}?
For too small Reynolds numbers towards larger $Pr$, say, $Re_{crit} = 50$,
 the distinction between
the bulk and the boundary layer is no longer meaningful;  
the bulk will no longer be driven to turbulence 
by a large scale velocity $U$.
Correspondingly, if the Nusselt number approaches 1 because of too small
$Pr$, the splitting of $\eps_\theta$ in
$\eps_{\theta, BL}$ and 
$\eps_{\theta, bulk}$ becomes meaningless. 
Finally, for  $Nu=1$, we no longer have thermal convection but pure 
thermal diffusion.

Therefore, we impose the restrictions 
$Re \aleq 50$ towards large $Pr$ and 
$Nu \ageq 1$ towards small $Pr$. The lines 
$Re=50$ and $Nu=1$ are
included in above phase diagram figure \ref{fig_sketch}.
Their analytical forms  
directly follow from the power laws
of table \ref{tab_scal_laws}; they are given 
in  table 
\ref{tab_bounds}.

Beyond these lines, in the shaded areas in figure \ref{fig_sketch}, 
the flow is viscosity dominated 
or thermal diffusivity dominated  
and the proposed power laws 
for $Re$ and $Nu$ (table \ref{tab_scal_laws}) no longer apply.

\subsection{Turbulence transition of the laminar boundary layer}

For very large $Ra$
the theory outlined here requires an  extension. 
It is based so far 
on the existence of a laminar  boundary layer flow
of Blasius type; its thickness therefore obeys $\lambda_u \sim L Re^{-1/2}$,
cf.\ 
section 39 of ref.\ \cite{ll87}.
The 
shear in this 
boundary layer is determined 
by the large scale velocity $U$ of the thermal roles in the RB cell
and the boundary layer width $\lambda_u$.
 We define
the corresponding  shear Reynolds number as 
\be
Re_{shear} = {U\lambda_u\over \nu} \sim \sqrt{Re}.
\label{s1}
\ee

%caption1
\begin{figure}[htb]
\setlength{\unitlength}{1.0cm}
\begin{picture}(11,12)
%%\put(0.5,12)
\put(0.5,0.5)
{\psfig{figure=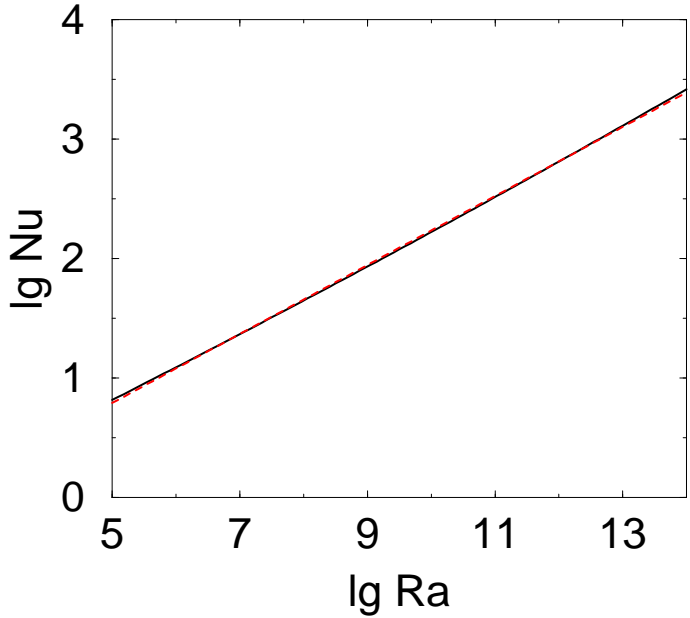,width=12cm,angle=-90}}
\end{picture}
\caption[]{
$Nu$ vs $Ra$ for fixed $Pr=1$ (characterizing helium)
according to the presented theory, i.e., equation (\ref{eq2733}) 
(solid line). Also shown is the practically indistinguishable
result  $Nu = 0.22 Ra^{0.289}$
of a linear regression of (\ref{eq2733}) (dashed line), 
which mimicks a power law with an exponent close to $2/7$. 
}
\label{nu_vs_ra_log_2733}
\end{figure}

The key issue now is that the laminar shear BL 
will become turbulent
for large enough $Re_{shear}$.
The details of the mechanism of this turbulence transition
is still under study but it seems to have {\it nonnormal-nonlinear} 
\cite{dra81,bob88,tre93,geb94,wal95,sch97,gro99}  features. What however is
agreed upon is that the 
shear Reynolds number at which  the turbulence sets in depends
on the kind  and strength
of the flow distortion. A typical value for the onset 
is \cite{ll87}
\be
Re_{shear,turb}  = 420.
\label{s2}
\ee
It will turn out that such high shear Reynolds numbers
can only be achieved in regimes II and IV;
in the regimes I and III the large Reynolds numbers
necessary for the breakdown of the laminar shear BL are not achieved.
With the information from table \ref{tab_scal_laws} we 
obtain the corresponding line in the $Ra-Pr$ parameter space indicating the 
laminar-turbulence onset range, namely 
\be
Pr_{turb} \sim Ra^{2/3}_{turb}
\label{s3}
\ee
in regimes $II_l$  and $IV_u$ (with different prefactors) and  
\be
Pr_{turb} \sim Ra^{1}_{turb}
\label{s4}
\ee
in regime $IV_l$.
The corresponding prefactors will be calculated later, but 
for clarity we already included the
characteristic lines which mark the onset to turbulence in the BL  
in the phase diagram figure \ref{fig_sketch}
as dashed lines. 
Above, the Rayleigh-Benard rolls are still laminar in the boundary layer,
below the boundary layer is turbulent.

What power laws for $Re$ and $Nu$ are to be expected in the regime
beyond the turbulence transition of the laminar BL?
One might argue that the destruction of the BL laminarity
means that
both the kinetic and the thermal
dissipation rates scale as in the turbulent bulk.
This implies that the scaling of $Re$ and $Nu$ 
should be the same as in the bulk dominated regime IV. In the phase 
diagram we called those regimes  $II_l^\prime$,
$IV_l^\prime$, and  
$IV_u^\prime$.

The same result is obtained by yet another argument:
In a turbulent thermal boundary layer it holds \cite{ll87,cha97}
\be
{L u_*\over \kappa} \sim Nu \log \left( {L u_*\over \kappa} \right).
\label{prandtl}
\ee
With logarithmic precision the typical velocity scale 
$u_*$ of the fluctuations in the BL is equal to the wind velocity 
$U$ and therefore eq.\ (\ref{prandtl}) implies
$
Nu\sim Re Pr. $
On the other hand, it still holds $\eps_u \sim \eps_{u,bulk}$
or
$
Nu Ra Pr^{-2}\sim Re^3$.
From 
these two relations 
one immediately obtains
the power laws 
(\ref{eq21}) and (\ref{eq22}), i.e., scaling as in regime $IV_l$
for all three primed regimes 
$II_l^\prime$,
$IV_l^\prime$, and 
$IV_u^\prime$. 

Still we feel that further study is necessary to obtain reliable 
insight about the dissipation rate scaling in turbulent boundary layers.
This might influence the scaling exponents in the primed regimes
$II_l^\prime$ and $IV_{l,u}^\prime$.

\section{Comparison with experiment: Scaling exponents}

\subsection{Nusselt number}
The first type of results of the theory which
we would like to compare with
experiments are the scaling exponents.
First, we focus on the Nusselt number.

For fixed
$Ra=10^6$ the $Nu$ number seems to increase up to $Pr \approx 7$,
 see figure \ref{fig_nu_vs_pr}. The fit to the experimental
data between $Pr=0.005$ and $Pr\approx 7$ gives
$Nu\sim Pr^{0.13\pm 0.02}$ in good agreement with the 
predicted exponent $1/8$ in the regime $I_l$. 
We  note, however, that the 
suggestion eq.\ (\ref{nu_chicago}) 
by Cioni et al.\ \cite{cio97} 
is also consistent with experiment
in the small $Pr$ regime.

 \begin{table}[htp]
 \begin{center}
 \begin{tabular}{|c|c|}
 \hline
          boundary between
       &   $Pr_{trans}$
\\
\hline
         $I_l-II_l$
       & $Pr_{trans} = 4.3 \cdot 10^{-8} Ra_{trans}^{2/3}$
\\
         $I_l-III_l$
       & $Pr_{trans} = 1.0 \cdot 10^{22} Ra_{trans}^{-2}$
\\
         $II_l-IV_l$
       & $Pr_{trans} = 9.7 \cdot 10^{10} Ra_{trans}^{-1}$
\\
         $III_l-IV_l$
       & $Pr_{trans} = 9.1 \cdot 10^{-12} Ra_{trans}^{1}$
\\
         $I_u-III_u$
       & $Pr_{trans} = 5.7 \cdot 10^{-33} Ra_{trans}^{3}$
\\
         $III_u-IV_u$
       & $Pr_{trans} = 4.8 \cdot 10^{-8} Ra_{trans}^{2/3}$
\\
\hline
         $I_l-I_u$
       & $Pr_{\lambda} = 2.0 Ra_{trans}^{0}$
\\
         $III_l-III_u$
       & $Pr_{\lambda} = 5.3 \cdot 10^{5} Ra_{trans}^{-1/2}$
\\
         $IV_l-IV_u$
       & $Pr_{\lambda} = 7.3 \cdot 10^{3} Ra_{trans}^{-1/3}$
\\
\hline
         $I_l$-(Re=50)
       & $Pr_{trans} = 6.7 \cdot 10^{-5}  Ra_{trans}^{2/3}$
\\         $I_u$-(Re=50)
       & $Pr_{trans} = 3.0 \cdot 10^{-3}  Ra_{trans}^{3/5}$
\\
         $I_l$-(Nu=1)
       & $Pr_{trans} = 3.5 \cdot 10^{4} Ra_{trans}^{-2}$
\\
         $II_l$-(Nu=1)
       & $Pr_{trans} = 1.2Ra_{trans}^{-1}$
\\
         $III_u$-(Re=50)
       & $Pr_{trans} = 2.9\cdot 10^{-5} Ra_{trans}^{2/3}$
\\
\hline
         $II_l$-(BL-turbul.)
       & $Pr_{trans} = 9.3 \cdot 10^{-12} Ra_{trans}^{2/3}$
\\
         $IV_l$-(BL-turbul.)
       & $Pr_{trans} = 3.4 \cdot 10^{-16} Ra_{trans}^{1}$
\\
         $IV_u$-(BL-turbul.)
       & $Pr_{trans} = 2.3 \cdot 10^{-11} Ra_{trans}^{2/3}$
\\
 \hline
 \end{tabular}
 \end{center}
\caption[]{
Boundaries between the various regimes I through IV,
towards the limiting regimes where $Nu =  1$ (small $Pr$)
and $Re = 50$ (large $Pr$),
and the nonnormal-nonlinear onset of shear turbulence 
(last three lines, see section II-F). 
}
\label{tab_bounds}
 \end{table}

%caption1
\begin{figure}[htb]
\setlength{\unitlength}{1.0cm}
\begin{picture}(11,12)
%%\put(0.5,12.)
\put(0.5,0.5)
{\psfig{figure=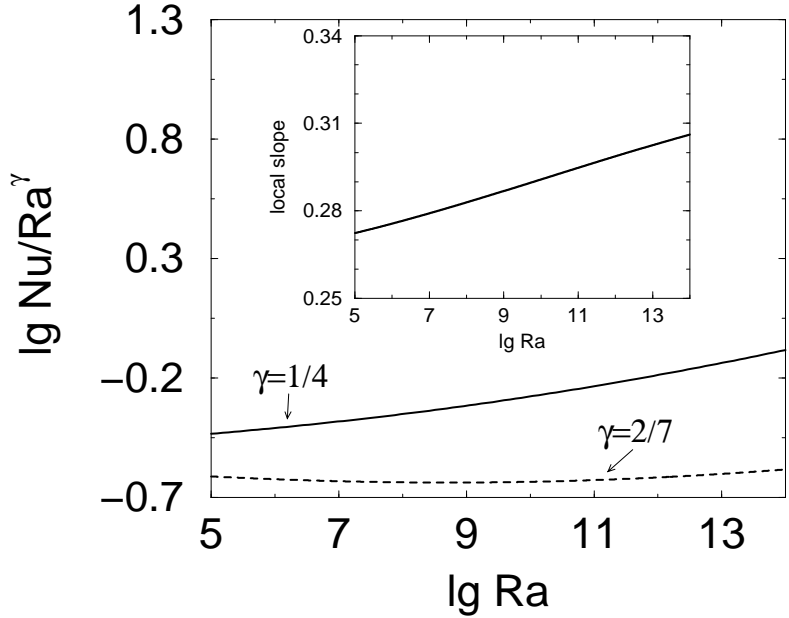,width=12cm,angle=-90}}
\end{picture}
\caption[]{
The $Nu$ number, as it follows from eq.\ (\ref{eq2733}, compensated
by two different power laws $Ra^{1/4}$ (solid, as suggest by the
present theory for the low $Ra$ regime) and $Ra^{2/7}$ (dashed).
The second is hardly distinguishable from a straight line, i.e.,
pure $2/7$ scaling. 
The inset shows the local slope following from eq.\ (\ref{eq2733}). 
}
\label{local_slope}
\end{figure}

%caption1
\begin{figure}[htb]
\setlength{\unitlength}{1.0cm}
\begin{picture}(11,12)
%%\put(0.5,12)
\put(0.5,0.5)
{\psfig{figure=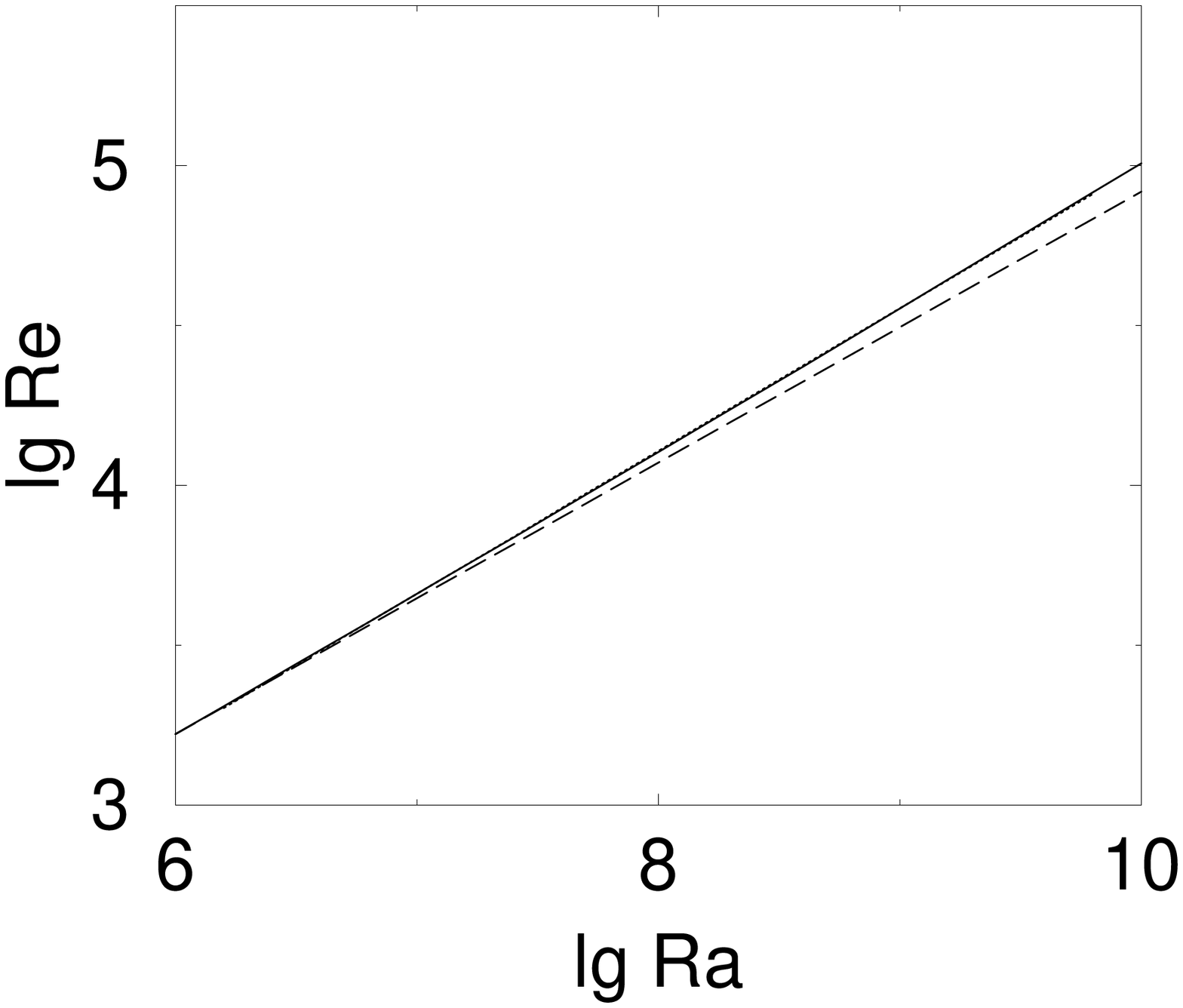,width=12cm,angle=-90}}
%%{\psfig{figure=/usr/people/lohse/rb/cioni/re_vs_ra_log_mod.ps,width=12cm,angle=-90}}
%{\psfig{figure=/vhe/therm2/mnt/wagner/lohse/rb/cioni/re_vs_ra_log_mod.ps,width=12cm,angle=-90}}
\end{picture}
\caption[]{
$Re$ vs $Ra$ for mercury, $Pr=0.025$.
The solid line shows the theoretical superposition
$0.59 Ra^{1/2} + 4.30 Ra^{2/5}$ which is very well fitted by
a straight line
$Re=3.5 Ra^{0.446}$ (dotted, practically indistinguishable
from the solid line; the fit interval is $Ra=10^6$ to $Ra=4\cdot 10^9$
as in experiment).
The dashed line presents Cioni et al.'s fit through their data
$Re \propto Ra^{0.424}$.
The prefactors cannot be compared because of the different definitions
of the Reynolds number in refs.\ \cite{cha97} and \cite{cio97}, see
section IV, first paragraph.
}
\label{fig_re_ra}
\end{figure}

The increase with $Pr$ seems to cease for $Pr$ between $1$ and $10$.
As stated above, other experimental data \cite{bel94} even suggest 
a {\it decrease} of $Nu$ with increasing $Pr$ in that regime. 
This is compatible with and explained by the present 
theory which gives 
a ($Ra$ number independent) transition from the 
$I_l$ regime with $Nu\sim Pr^{1/8}$ to $I_u$ with $Nu\sim Pr^{-1/12}$,
both for fixed $Ra$. 

As there is even a controversy in literature whether $Nu$ 
for water with $Pr=6.6$ 
or $Nu$ for helium gas with $Pr=0.7$
 is larger, it is hard to say
where exactly the transition from $I_l$ to $I_u$ takes place.
According to Cioni et al.\ \cite{cio97}  it is fair to say that within experimental
accuracy $Nu (Pr=6.6) = Nu(Pr=0.7)$. We adopt this point of view
and use it to calculate the transitional $Pr$ number $Pr_\lambda^{I_l-I_u}$
to be about $2$. This experimental
information thus defines the line $\lambda_u 
=\lambda_\theta$ in the phase space and separates the lower subregime
$I_l$
with $\lambda_u < \lambda_\theta$ from the upper subregime $I_u$ with 
with $\lambda_u > \lambda_\theta$.

Next, we compare the predicted scaling exponent $\gamma$
of $Nu$ vs $Ra$, which in regime I is according to the presented theory 
the same for the lower and the upper subregime. 
For small  $Pr$ 
(mercury, sodium)
and (relatively)
small $Ra$ the theoretically obtained value $\gamma = 1/4$ for the scaling exponent
of $Nu$ vs $Ra$ has been  measured in several experiments:
$\gamma =0.247 $ in \cite{ros69},
$\gamma =0.26\pm 0.02 $ in \cite{cio97}, and
$\gamma =0.25 $ in \cite{hor98}.

Also regime $II_l$ (intermediate $Ra$, low $Pr$)
with $\gamma = 1/5$ seems to have been
observed by Cioni et al.\ \cite{cio97}.

For larger 
$Pr$ 
(helium, water)
the measured scaling exponent is larger, $\gamma \approx 2/7$,
see table \ref{tabneu}.
Here we argue
that this results from the superposition of the
scaling in regime I, with those in the regimes $IV_u$ 
and $III_u$.
To substantiate this,
we plot the expected $Nu$ vs $Ra$ dependence for 
$Pr=1$,
\be
Nu = 0.27 Ra^{1/4} + 0.038 Ra^{1/3},
\label{eq2733}
\ee
in figure \ref{nu_vs_ra_log_2733}. Here we have
already made use of the prefactors from table \ref{tab_scal_laws},
which will be calculated in the next section.
Now we fit eq.\ (\ref{eq2733}) 
with {\it one} power law in as large a  regime 
as $10^5 \le Ra \le 10^{14}$. 
This fit which is nearly indistinguishable from the superposition
(\ref{eq2733}) reads
\be
Nu = 0.22 Ra^{0.289}.
\label{eq27allein}
\ee
The power law exponent is very close to $2/7=0.286$ 
and definitely consistent with the experimental data of table
\ref{tabneu}.
Changing the fit regime of course changes the exponent
of the power law (\ref{eq27allein}). E.g., for a 
linear regression
in the regime $10^6 < Ra < 10^{11}$ which is typical for
many experiments one obtains 
$Nu = 0.24 Ra^{0.285}$, i.e., an exponent which is even closer to
$2/7$. 

By plotting compensated plots or local slopes $d \log_{10} Nu /d \log_{10} Ra$
as done in figure \ref{local_slope} for eq.\ (\ref{eq2733})
one 
may be able to get hints that there is no pure power law.
Note, that on first sight a compensation with $Ra^{2/7}$ may erroneously
even 
be considered as ``better''.

Chavanne et al.\ \cite{cha97}
find hints for a transition to a regime with a visibly larger
scaling exponent. According to our theory
this could be regime $III_u$ or $IV_l$ or $IV_l^\prime$ or, most likely,
a mixture of all of them. 
No clean scaling exponent could hitherto be determined experimentally. 
One reason
is that in that regime both $Ra$ and $Pr$ change.
We will discuss the possible nature of this transition below. 
Also figure 3 of ref.\ \cite{sig94} suggests such a transition.

We now turn to the large $Pr$ regime. 
There are very few data for large $Pr \gg 1$. Recently,
Ashkenazi and Steinberg
\cite{ash99} performed
convection experiments with SF$_6$ close to
its critical point. In these experiments both $Ra$ and $Pr$ 
change considerably at the same time.
To what degree 
 the RB convection is still
Boussinesq close to the critical point is extensively
discussed in ref.\ \cite{ash99}. 

Ashkenazi and Steinberg obtain 
$Nu=0.22 Ra^{0.3\pm 0.03} Pr^{-0.2 \pm 0.04}$
in $10^{9} \le Ra < 10^{14}$ and 
$1 \le Pr \le 93$.
Based on the phase diagram \ref{fig_sketch} we judge
that for these ranges of the $Ra$ and $Pr$ numbers we should be 
in regimes $I_u$ and $III_u$. 
From table \ref{tab_scal_laws} we see that the
$Ra$ exponent of $Nu$ is $1/4$ and $3/7$, respectively. 
The measured exponent of $0.3\pm 0.03$ in between is
consistent with this. 
The $Pr$ exponent of $Nu$ is expected to be in between
$-1/12$ and $-1/7$, slightly smaller (modulus-wise) than
the value of $-0.2\pm 0.04$ reported in \ \cite{ash99}.

The results 
of this subsection 
clearly demonstrate
the importance of superimposing the power laws of adjacent 
phase space regimes. This can
really mimick different scaling behavior, as demonstrated in
figures 
\ref{nu_vs_ra_log_2733} and \ref{fig_re_ra}. This characteristic feature 
holds because
the power law exponents of the neighboring regimes are rather similar. They 
will commonly
show up if data in a crossover range are examined. We emphasize 
that these crossover
ranges appear to be rather extended, reaching well into the corresponding
regimes, due to the 
 small differences of the scaling exponents.  

Therefore, rather than writing pure power laws, one should allow
for superpositions. Table \ref{tab_scal_laws} suggests 
 \be
 Nu \sim Ra^{1/4} Pr^{1/8} \cdot \left( 1+  
 \left\{\begin{array}{cc}
  c_{III_u}& Ra^{5/28} Pr^{-15/56} +\cdots
  \\
  c_{IV_u}& Ra^{1/12} Pr^{-1/8} +\cdots
  \\
  c_{IV_l}& Ra^{1/4} Pr^{3/8} +\cdots
  \\ 
  c_{II_l}&Ra^{-1/20}Pr^{3/40}+\cdots
   \end{array}\right.
   \right)
\label{eq35}
\ee
and 
\be
 Re \sim  Ra^{1/2} Pr^{-3/4} \cdot \left( 1+
 \left\{\begin{array}{ll}
  c^\prime_{III_u}& Ra^{1/14} Pr^{-3/28} +\cdots
  \\
  c^\prime_{IV_u}& Ra^{-1/18} Pr^{1/12} +\cdots
  \\
  c^\prime_{IV_l}&  Pr^{1/4} +\cdots
  \\ 
  c^\prime_{II_l}&Ra^{-1/10}Pr^{3/20} +\cdots
   \end{array}\right.
   \right),  
\label{eq36}
\ee
respectively. Here, we have separated the exponents of  regime $I_l$. 
Which of these corrections and how many are  to be taken depends on the 
$Pr$ number
and on the aspect ratio.
{\it Locally}, 
i.e., for a limited $Ra$ range, the suggested $Nu$ vs $Ra$ power law exponents
$\gamma = 2/7$ \cite{cas89,shr90} (cf.\ figure 2 of ref.\ \cite{cha97}
or figure 3 of \cite{sig94}), or
$\gamma = 5/19$ \cite{yak92} can still be considered as an appropriate
representation of the experimental data. {\it Globally},
for larger $Ra$ intervals, however, we
claim that eq.\ (\ref{eq35}) is a better description.

\begin{figure}[htb]
\setlength{\unitlength}{1.0cm}
\begin{picture}(11,12)
%%\put(0.5,12)
\put(0.5,0.5)
{\epsfig{figure=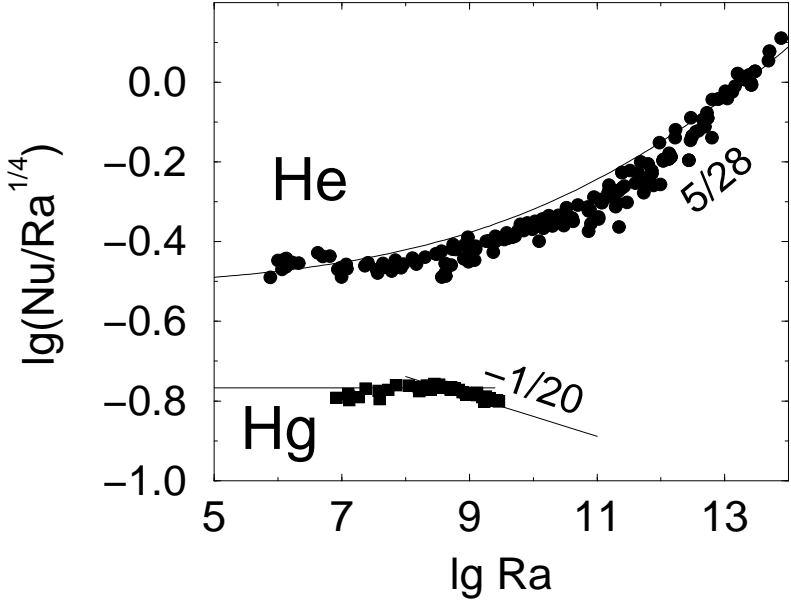,width=12cm,angle=-90}}
%%{\epsfig{figure=/usr/people/lohse/rb/chavanne/nu_vs_ra_log1.eps,width=12cm,angle=-90}}
%{\epsfig{figure=/vhe/therm2/mnt/wagner/lohse/rb/chavanne/nu_vs_ra_log1.eps,width=12cm,angle=-90}}
%
%{\psfig{figure=[detlef.rb.slides]NU_VS_RA_COMP.PS,width=12cm,angle=-90}}
\end{picture}
\caption[]{
Compensated $Nu$ vs $Ra$ data for $Pr$ from about 1 through about 20
(helium, upper, data taken
from ref.\ \cite{cha97}; the higher $Ra$ experiments also have higher $Pr$-number)
and $Pr=0.025$ (mercury, lower, taken from ref.\ \cite{cio97}).
Also shown are the calculated exponents in the large
$Ra$ regimes.
We also drew the theoretical curve  
$Nu=0.33Ra^{1/4}Pr^{-1/12} + 3.43 \cdot 10^{-3}
 Ra^{3/7}Pr^{-1/7}$ with fixed $Pr=3$ to demonstrate that it roughly 
describes the data. If the expected 
$Pr$ number dependence is considered, the agreement becomes even better.
}
\label{fig_nu_vs_ra}
\end{figure}

In previous publications $Nu$, compensated by the expected
scaling $Ra^{2/7}$, was plotted against $Ra$ in a log-log plot,
see figure 2 of ref.\ \cite{cha97}.  From that plot one realizes
that the 2/7-scaling  is slightly too steep between $Ra=10^6$ and
$Ra=10^8$ and  not steep enough beyond the crossover
at $Ra=10^{11}$. The analogously compensated plot with the expected scaling 
(\ref{eq29}) is shown in figure 
\ref{fig_nu_vs_ra}. Now (for $Pr\approx 1$)
one obtains a horizontal line up to
$Ra\approx 10^9$, showing that eq.\ (\ref{eq29}) nicely agrees with the
data. However, beyond $Ra\approx 10^9$ one observes  deviations.
We suggest that these corrections originate from the different scaling
in the neighboring regime $III_u$.
The reason that it is regime $III_u$ (and not regime $IV_u$) is that 
in the Chavanne et al. experiments
the large $Ra$ measurements also have large $Pr$; the trajectory in 
control parameter space $Ra-Pr$ is not a straight
line. At $Ra=10^{10}$ one typically   
has $Pr\approx 1$, but at $Ra=10^{14}$ Chavannve et al.\
typically have $Pr\approx
10-20$. 
The power law exponent $5/28$, following from table
\ref{tab_scal_laws}, 
is consistent with 
the 
experimental data 
for large $Ra$, see figure \ref{fig_nu_vs_ra}.

For the mercury data of Cioni et al. \cite{cio97} we do not have
such a complication as $Pr=0.025$ is roughly constant for all chosen $Ra$.
As plotted in figure \ref{fig_nu_vs_ra}, lower curve, we observe a 
straight line
up to about $2\cdot 10^8$ and then a decay,
signaling contributions from regime $II_l$.
The power law exponent $-1/20$ of the correction term in eq.\ (\ref{eq35})
is consistent with the data shown in figure \ref{fig_nu_vs_ra}.

\begin{figure}[htb]
\setlength{\unitlength}{1.0cm}
\begin{picture}(11,12)
%%\put(0.5,12)
\put(0.5,0.5)
{\epsfig{figure=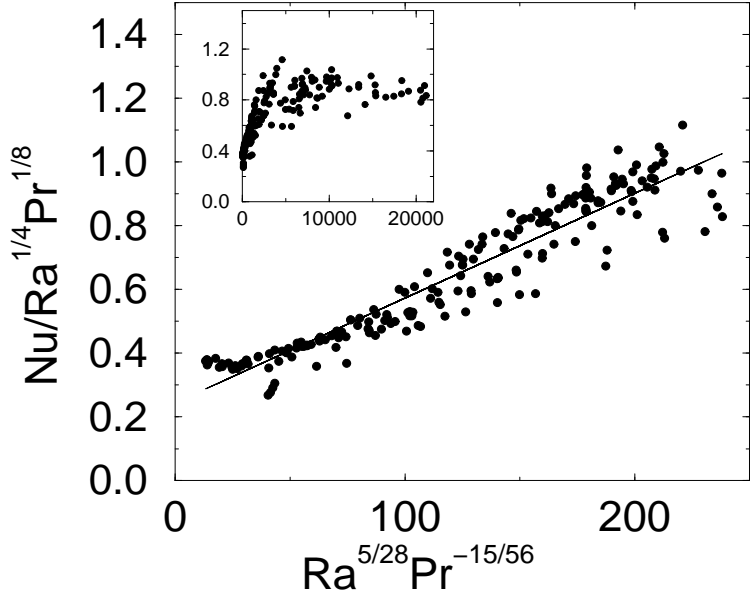,width=12cm,angle=-90}}
\end{picture}
\caption[]{
Same data as in the previous figure, but now in a 
linear plot
$Nu/(Ra^{1/4}Pr^{1/8})$ vs $Ra^{5/28}Pr^{-15/56}$,
revealing the quality of the superposition eq.\ (\ref{eq35}), 
high $Pr$, regime $III_u$. 
The linear fit (straight line) gives
$Nu/(Ra^{1/4} Pr^{1/8}) = 0.24 + 3.3\cdot 10^{-3} Ra^{5/28}Pr^{-15/56}$. 
The data points are taken from ref.\ \cite{cha97}, with kind permission
by the authors. Only data points with $Ra>10^6$ are 
considered. Note that in this plot $Pr$ varies as much as $0.6 < Pr < 100$.
-- The inset shows 
$Nu/(Ra^{1/4}Pr^{1/8})$ vs $Ra^{1/4}Pr^{3/8}$; this variable had to 
be used if regime $IV_l$ contributes the most relevant correction.
The data do not fall on a straight line. A similar failure results 
with the regime $IV_u$-compensated variable $Ra^{1/12}Pr^{-1/8}$.
}
\label{fig_leo}
\end{figure}

A more stringent  way to test the
superpositions of type (\ref{eq35}) 
is to make a {\it linear} plot $Nu/(Ra^{1/4}Pr^{1/8})$ 
vs $Ra^{5/28}Pr^{-15/56}$ or vs $Ra^{1/12}Pr^{-1/8}$
or vs $Ra^{1/4}Pr^{3/8}$, etc, depending on 
which neighboring regime the corrections originate. 
This is done 
in figure \ref{fig_leo}, assuming, as argued above, that
the most relevant corrections originate from regime $III_u$. 
If the theory is correct, the data points
must fall on a straight line. 
Indeed, they do so with satisfying precision. 

Note that this kind of linear plot is very sensitive to what
combinations of $Ra$ and $Pr$ are chosen as $x$ and $y$ axes. 
E.g., plotting
$Nu/(Ra^{1/4}Pr^{1/8})$ vs $Ra^{1/4}Pr^{3/8}$ (the subleading correction
characterizing regime $IV_l$) does not lead to a straight line
at all, see the inset of figure 
\ref{fig_leo}.
Clearly the variable $Ra^{5/28}Pr^{-15/56}$ on the abszissa is superior,
adding confirmation that the Chavanne et al.\ \cite{cha97} large $Ra$ 
experiments represent the physics of regime $III_u$.

\subsection{Reynolds number}
We now consider  the experimental values for the scaling exponents
$\alpha$ of the Reynolds number vs the Rayleigh number.
For $Pr\approx 7$ (water) 
Xin et al.\ \cite{xin96} find $\alpha = 0.50 \pm 0.01$
and Qiu and Xia \cite{qiu98} find  $\alpha = 0.50 \pm 0.02$. Both
experiments where done in the Ra interval between $2\cdot 10^8$ and 
$2\cdot 10^{10}$, i.e., in regime $I_u$, where exactly this power
law exponent $\alpha = 1/2$ is expected.
For $Pr\approx 1$ both Castaing et al.\ \cite{cas89} and
Chavanne et al.\ \cite{cha97} find $\alpha= 0.49$ for all $Ra$ which
suggests that possibly the regimes $I_l$ and $IV_l$
 are seen where this value is
predicted, or also
regime $III_u$, where the exponent is only slightly higher ($4/7$).
 For $Pr = 0.025$ (mercury) 
Cioni et al.\ \cite{cio97} find $\alpha = 0.424$
from a fit to all available $Ra$. This value is in between the derived
values $\alpha = 1/2$ in regime $I_l$ and $\alpha = 2/5$ in 
regime $II_l$.

We compare 
this experimental finding
$Re \propto  Ra^{0.424}$
(based on a fit to the data
in the range up to $Ra=4\cdot 10^9$ \cite{cio97}) 
with the $I_l$-$II_l$ superposition according to table \ref{tab_scal_laws} 
\be
Re= 0.59 Ra^{1/2} + 4.30 Ra^{2/5},
\label{superpos_re}
\ee
 cf.\ 
fig.\ \ref{fig_re_ra}. In this relatively
short $Ra$ interval the theoretical
superposition (\ref{superpos_re})
 is again hardly distinguishable from its straight line fit
\be
Re=3.5 Ra^{0.446}
\label{slf_re}
\ee
whose exponent reasonably well agrees with
the measured one. 
{\footnote{The absolute values of the Reynolds numbers
cannot be compared here, as the Reynolds number definitions
in refs.\ \cite{cio97} and \cite{cha97} cannot be transferred 
into each other, see the first paragraph in section IV.}}

\begin{figure}[p]
\setlength{\unitlength}{1.0cm}
\begin{picture}(18,18)

%%\put(3.75,9.)
\put(4.0,0.5)
{\psfig{figure=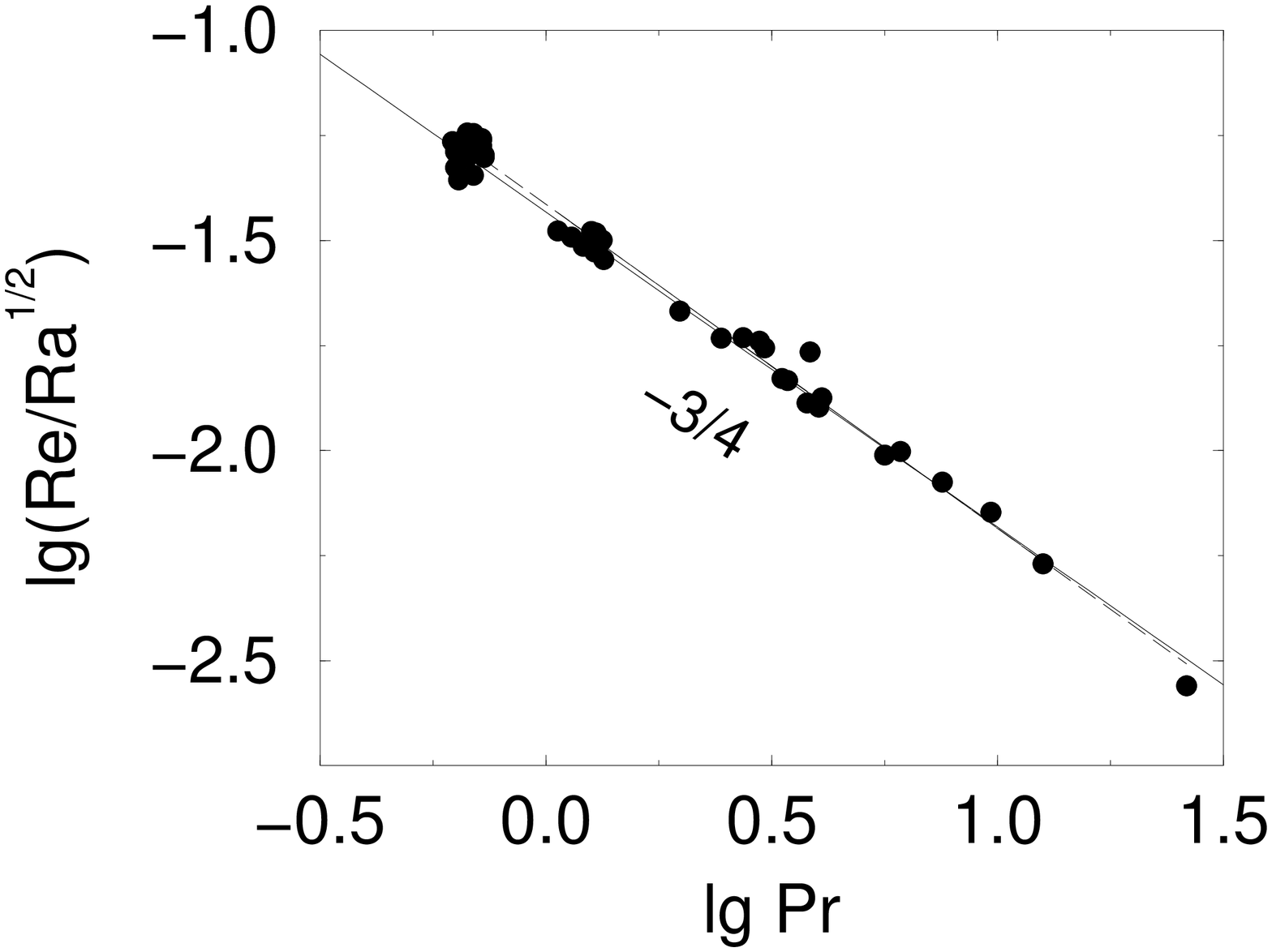,width=8.8cm,angle=-90}}
%%{\psfig{figure=/usr/people/lohse/rb/chavanne/re_vs_pr_log_he.eps,width=11cm,angle=-90}}

%%\put(4.75,18)
\put(4.75,9.0)
{\psfig{figure=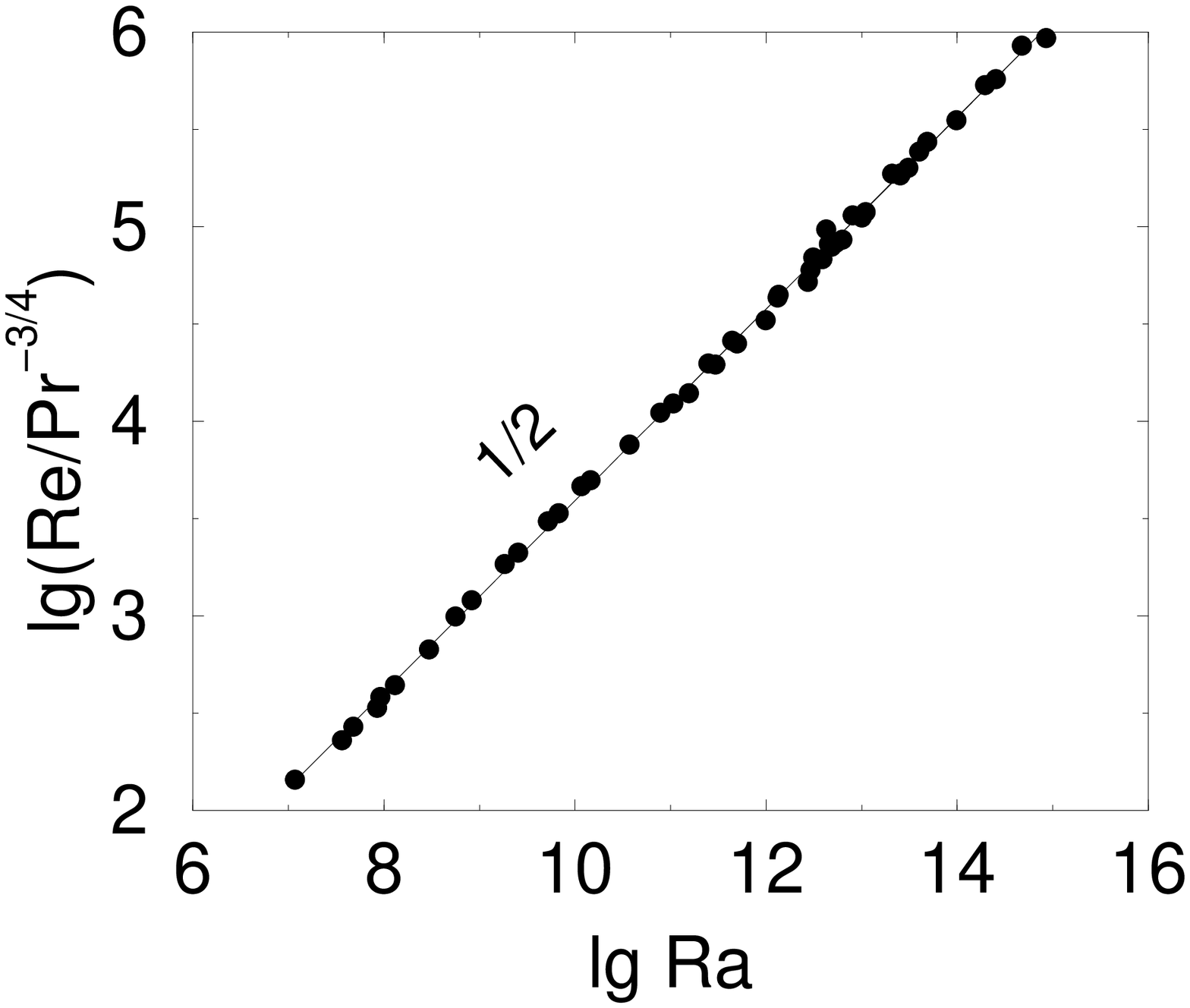,width=8.5cm,angle=-90}}
%%{\psfig{figure=/usr/people/lohse/rb/chavanne/re_vs_ra_log_he.eps,width=9.5cm,angle=-90}}
%{\psfig{figure=/vhe/therm2/mnt/wagner/lohse/rb/chavanne/re_vs_pr_log_he.eps,width=11cm,angle=-90}}
%\put(3.75,10.)
%{\psfig{figure=/vhe/therm2/mnt/wagner/lohse/rb/chavanne/re_vs_ra_log_he.eps,width=9.5cm,angle=-90}}
\put(1.0,5.5){\Large{(b)}}
\put(1.0,13.5){\Large{(a)}}
\end{picture}
\caption[]{
The figure presents data for the Reynolds number 
as a function of $Ra$ and $Pr$ 
from ref.\ \cite{cha97}, with very kind
permission of the authors. \\
a)
$Re/Pr^{-3/4}$ vs $Ra$. The expected slope is $1/2$, the linear regression
fit (solid line) gives $0.492\pm 0.002$.\\
b)
$Re/Ra^{1/2}$ vs $Pr$. The expected slope is $-3/4$, the linear regression
fit (dashed line) gives $-0.77\pm 0.01$. 
The agreement of the prefactor is also excellent. 
According to theory, following table \ref{tab_scal_laws} it is 
$\log_{10}(Re/Ra^{1/2}) = -1.432 -(3/4)\log_{10} Pr$ (solid line, hardly
distinguishable from the dashed one); 
the fit value for the prefactor from linear regression 
is $-1.413\pm 0.005$. 
}
\label{fig_re_he}
\end{figure}

Moreover, also the 
theoretically 
obtained $Pr$ number dependence of $Re$ very nicely
agrees with available experimental information: Chavanne
et al.\ \cite{cha97} did
experiments with (slightly) varying $Pr$. They then plotted
$RePr^{0.72}$ vs $Ra$ and obtained the law
$Re Pr^{0.72} = 0.0374 Ra^{1/2}$. 
The  exponent $0.72$ of the Prandtl number was
determined by minimizing the scattering of points around a straight
line in the log-log plot.
 It very well agrees with the calculated
$Pr$ scaling exponent 3/4 in eq.\ (\ref{eq30}) (regime $I_l$). A possible
reason for the slight difference between theory and data fit
is that part of the experimentally realized $Ra$ and $Pr$ already 
belong to regimes $III_u$ and $IV_u$,
where according to eqs.\ 
(\ref{eq22}) and
(\ref{eq22u})
 the expected $Pr$ scaling exponent
is $6/7$ and $2/3$, respectively. However, the deviation
is clearly within the experimental uncertainty. 
In figure \ref{fig_re_he} we replot the experimental Reynolds number
data of Chavanne
et al.\ \cite{cha97}. In fig.\ \ref{fig_re_he}a we show 
$Re/Pr^{-3/4}$ vs $Ra$. The fit gives a $Ra$-exponent $0.492\pm 0.002$
in very good agreement with the theory's exponent $1/2$. 
In fig.\ \ref{fig_re_he}b we display $Re/Ra^{1/2}$ vs $Pr$. The data fit
results in a $Pr$-exponent $-0.77\pm 0.01$, also in excellent agreement
with the theoretical expectation 
which is $-3/4$ in $I_l$ and $-5/6$ in $I_u$.

The large $Ra$ values in both figures will turn out to belong
already to the regimes $III_u$ and $IV_u$, where the $Pr$-scaling
is similar. The available range is too small to perform a more
detailed comparison.

The only large $Pr$ data available are again those by Ashkenazi and
Steinberg \cite{ash99}. They obtain 
$Re=2.6 Ra^{0.43\pm 0.02} Pr^{-0.75 \pm 0.02}$
in $10^{12} \le Ra < 3\cdot 10^{14}$ and 
$27 \le Pr \le 190$.
The expected $Ra$ exponent of $Re$ is between $1/2$ and
$4/7$, distinctly larger than the measured one of $0.43 \pm 0.02$. 
Similarly, the theoretically expected $Pr$ exponent of $Re$ is between
$-5/6$ and $-6/7$, also larger than the measured
exponent $-0.75 \pm 0.02$. We have no explanation.

%caption1

%caption1

%caption1

\section{Prefactors}
\subsection{Experimental input to determine the prefactors}
To obtain the values of the prefactors in the power laws 
within the presented scaling
theory, we need further 
input from experiment. 
But what data to choose? As pointed out in the introduction,
a huge variety of data is around, often disagreeing with each other 
even in the scaling exponents, not to speak of prefactors. 
Those often vary by as much as $50\%$ from experiment to experiment,
even for cells with the same aspect ratio.
Another reason which makes the adoption to one experiment
and the comparison to others difficult is that 
different definitions are used for the Reynolds numbers.
E.g., Cioni et al.\ \cite{cio97} define the Reynolds number as
$Re= 4L^2f_p/\nu$, where $f_p$ is a distinguished frequency
at the small frequency edge in the temperature spectrum.
Chavanne et al.\ \cite{cha97} define $Re= \omega_0 d L/\nu$,
where $\omega_0$ is a typical frequency in the cross correlation
spectrum of two temperature signals, measured at a vertical distance
of $d=2.3$ mm and $2$ cm off the axis of the cell. 

In spite of these difficulties, we decided to calculate the 
prefactors of the suggested scaling laws
by employing the following choice for the input information as a 
reasonable, realistic example. Our reasons are first, to be able
to draw a phase diagram with more or less realistic values.
Second, to stress the importance of the prefactors. 
But we caution the reader that our input choice 
is somehow arbitrary; other possibilities 
can equally well be rationalized, sometimes
shifting the various regime boundaries considerably.

Above, as an input from experiment, we had already chosen
$Pr_\lambda =2$ as the Prandtl number for which $Nu$ is maximal (for fixed
$Ra = 10^{6}$, cf.\ 
fig.\ \ref{fig_nu_vs_pr}). 
In addition, we will use the following 
experimental information:
\begin{enumerate}
\item
The observed transition Rayleigh number for the transition from 
regime $I_l$ to regime $III_l$, $Ra_{trans}=10^{11}$ at $Pr=1$ \cite{cha97};
\item
the observed transition Rayleigh number for the transition from 
regime $I_l$ to regime $II_l$, $Ra_{trans}=
4.5 \cdot 10^{8}$ at $Pr=0.025$ \cite{cio97};
\item 
the experimental values, taken from ref.\ \cite{cha97}, for 
the Reynolds and the Nusselt number at the middle point
$(Ra_M,Pr_M)$ in the phase diagram figure \ref{fig_sketch};
\item 
and the prefactor 0.0372 of the scaling law 
$Re=0.0372 Ra^{1/2} Pr^{-3/4}$ measured in regime $I_l$ \cite{cha97}.
\end{enumerate}

Information (1)
specifies
the prefactor of the rhs of eq.\ (\ref{eq32})
(which we call $c_{I_l-III_l}$) to be
$c_{I_l-III_l} = Pr_{trans}^{I_l-III_l} /Ra_{trans}^{-2} = 1.0 \cdot 10^{22}$.
Information (2) gives 
the prefactor of the rhs of eq.\ (\ref{eq31})
(which we call $c_{I_l-II_l}$) to be
$c_{I_l-II_l} = Pr_{trans}^{I_l-II_l} /Ra_{trans}^{2/3} = 4.3 \cdot 10^{-8}$.
The two curves (\ref{eq31}) and (\ref{eq32}) cross at
\be
(Ra_M, Pr_M) = (1.03 \cdot 10^{11}, 0.94).
\label{n1}
\ee
This middle point 
$(Ra_M, Pr_M)$
is defined by the conditions
$\eps_{u,BL} = \eps_{u,bulk} = \eps_u /2$ and 
$\eps_{\theta,BL} = \eps_{\theta,bulk} = \eps_\theta /2$.
Equation (\ref{n1})
specifies the prefactors of the rhs of eqs.\ (\ref{eq33}) and
(\ref{eq34}) to be
$c_{III_l-IV_l} = 9.1 \cdot 10^{-12}$ and 
$c_{II_l-IV_l} = 9.7 \cdot 10^{10}$, respectively. 
We see that $Pr^{I_l-I_u}_{\lambda}=2 > Pr_M = 0.94$ so that there
is a regime $III_l$, but no regime $II_u$, as already anticipated.
The line $Pr_{\lambda}^{I_l - I_u}=2 $ hits the boundary
between $I$ and $III$ at 
\be
(Ra_{M'}, Pr_{M'}) = (7.1 \cdot 10^{10} , 2.0),
\label{mprime}
\ee
which fixes the prefactors of eqs.\
(\ref{eq32u}) and (\ref{tll3}) to be $5.7 \cdot 10^{-33}$ 
and $5.3 \cdot 10^{5}$, respectively. Correspondingly, one obtains
the prefactors of eqs.\ 
(\ref{eq33u}) and (\ref{tll4})
to be 
$4.8\cdot 10^{-8}$ and 
$7.3\cdot 10^{3}$, respectively. 
These are the data on
which the phase diagram figure \ref{fig_sketch} is based;
they are summarized in table \ref{tab_bounds}.

Apart from regime $III_l$ all regimes turned out to have
at least one decade of extension both in $Ra$ and in $Pr$ 
and should therefore in principle be visible.
However,
we should always expect one or more subleading corrections.
Regime $III_l$ will clearly not be detectable. 

We again stress how dependent this phase diagram drawn in 
figure \ref{fig_sketch}
is on the choice
of experimental information. E.g., if we had adopted Glazier et al.'s 
\cite{gla99} point of view that there is no transition 
towards a steeper $Ra$ dependence of $Nu$ at least up to 
$Ra = 8\cdot 10^{10}$,
regimes $II$ and $IV$ would have shifted further to the right or would
even not exist at all. But as shown in 
figure \ref{nu_vs_ra_log_2733},
apparent 
smooth scaling behavior does {\it not} allow to exclude a transition.

Making use now of the experimental information (3) and (4) we can calculate
the prefactors in the power laws for $Nu$ and $Re$. 
From figs.\ 2 and 3 of Chavanne et al.'s work \cite{cha97} we can  
extract the Reynolds
and Nusselt numbers at the middle point
$(Ra_M, Pr_M)$
of the phase diagram
which touches all four regimes,
namely
\be
Re_M = 1.20 \cdot 10^4,\qquad Nu_M = 2.78 \cdot 10^2,
\label{n2}
\ee
which is 
information (3) above.
The definition of the middle point 
$(Ra_M, Pr_M)$,
i.e., the conditions
$\eps_{u,BL} = \eps_{u,bulk} = \eps_u /2$ and 
$\eps_{\theta,BL} = \eps_{\theta,bulk} = \eps_\theta /2$, 
allows to calculate the prefactors
$c_{\eps_u ,bulk}$, 
$c_{\eps_\theta ,bulk}$, and
$c_{\eps_u ,BL}$  
on the rhs of eqs.\ (\ref{eq17}), (\ref{eq18}), and (\ref{eq19}),
respectively. One obtains
\be
c_{\eps_u , bulk} =
{Nu_M Ra_M \over 2 Pr^2_M Re_M^3} = 9.38,
\label{n3}
\ee
\be
c_{\eps_\theta , bulk} =
{Nu_M \over 2 Pr_M Re_M} = 0.0123,
\label{n4}
\ee
\be
c_{\eps_u , BL} =
{Nu_M Ra_M \over 2 Pr^2_M Re_M^{5/2}} = 1028. 
\label{n5}
\ee
Finally, 
the prefactor $c_{Nu}$ on the rhs of
relation (\ref{eq26}) is adopted
to Chavanne et al.'s \cite{cha97}
experimentally determined prefactor (see figure 3 of that paper)
in the relation (\ref{eq30}), i.e., 
$Re= 0.0372 Ra^{1/2} Pr^{-3/4}$ valid throughout regime $I_l$
(information (4) above). 
In that regime $\eps_{u,BL} = \eps_u$;
with eqs.\
(\ref{eq11}),
(\ref{eq19}), and
(\ref{n5}) 
we get
\be
c_{Nu} =
(3.72 \cdot 10^{-2})^2 \cdot c_{\eps_u , BL} = 1.42.
\label{n6}
\ee
The prefactors referring to the upper halve of the phase diagram are
calculated from the matching conditions for  $Re$ and $Nu$ on the
$\lambda_u = \lambda_\theta$ line.
With eqs.\ (\ref{n3}) -- (\ref{n6})
and the matching conditions
now {\it all} prefactors of the power laws in the four different
regimes are determined. We have summarized all these power laws in table
\ref{tab_scal_laws}.

Note that from  the condition $\lambda_u = \lambda_\theta $ 
for $Pr=2$ the prefactor of eq.\ (\ref{eq14}) is also obtained.
It is $0.25$, i.e., $\lambda_u = 0.25 L Re^{-1/2}$. 
Also the prefactors to the $Nu= 1$- and the $Re\le  50$-borders of validity
and
the crossover lines to turbulence of the laminar BL flow automatically follow,
and are included into table \ref{tab_scal_laws}.

\subsection{Comparison of the evaluated prefactors to experiment}

We now would like to compare the absolute agreement of
the power laws summarized in table \ref{tab_scal_laws},
whose prefactors result from an adoption to above four
pieces of experimental information, 
with {\it further} experimental
data. 
We first focus on the Chavanne et al.'s\ \cite{cha97} RB measurements in
helium gas. From the previous section we know that in these experiments
due to the large $Pr$ at large $Ra$ it is mainly regime $III_u$
which causes additional contributions to regime $I_l$. 
Therefore, we have plotted the superposition  
$Nu=0.33Ra^{1/4}Pr^{-1/12} + 3.43 \cdot 10^{-3}
 Ra^{3/7}Pr^{-1/7}$ with $Pr=3$ into figure \ref{fig_nu_vs_ra}.
We again stress that in
the experiments $Pr$ is not
constant at all.
Nevertheless, the data are satisfactorily described. Note that the 
solid curve in figure  \ref{fig_nu_vs_ra} is no fit!

A much better way to check whether the obtained  prefactors of the
theory agree with the measured ones is to do a linear regression
of the straight line as offered in figure \ref{fig_leo}. 
Such a straight line fit gives 
$Nu/(Ra^{1/4}Pr^{1/8}) = 0.24 + 3.3\cdot 10^{-3} Ra^{5/28}Pr^{-15/56}$. 
The found prefactors are in  good agreement
with the expectation
$Nu/(Ra^{1/4}Pr^{1/8}) = 0.27  +
3.43 \cdot 10^{-3} Ra^{5/28}Pr^{-15/56}$
from table \ref{tab_scal_laws}.

We also compare the theoretical
prefactors of the Reynolds number scaling in regime $I_l$ with 
experiment. 
As the theoretical prefactors have been adopted
to the experimental $Re/Pr^{-3/4}$ vs $Ra$ power law, it only makes
sense to check the prefactors in
$Re/Ra^{1/2}$ vs $Pr$. From table \ref{tab_scal_laws}
the expected slope is $-3/4$
and the expected prefactor $0.037$.
Linear regression gives a slope of $-0.77\pm 0.01$ and
a prefactor of $10^{-1.413\pm 0.005} = 0.0386$, cf.\ figure 
\ref{fig_re_he}.

Next, we compare the experimental $Nu$ vs $Ra$ scaling
for mercury with $Pr=0.025$ with theory.
The measured relations
$Nu = (0.140\pm 0.005) Ra^{0.26\pm 0.02}$ \cite{cio97},
$Nu = 0.147  Ra^{0.257}$ \cite{ros69}, and 
$Nu = 0.155 Ra^{0.27}$ \cite{tak96}
are all in reasonable agreement with the regime $I_l$ expectation 
$Nu= 0.17 Ra^{1/4}$ from table \ref{tab_scal_laws}.
The same holds for a comparison in regime $II_l$:
The reported experimental fit  is 
$Nu = 0.44\pm 0.015 Ra^{0.20\pm 0.02}$ \cite{cio97},
theory gives 
$Nu= 0.46 Ra^{1/5}$, again, remarkable agreement of both the power
law exponent {\it and} the prefactor. Remember that the
only experimental input from this experiment into the theory
is $Ra_{trans}^{I_l-II_l} = 4.5\cdot 10^8$.

Let us also check the prefactors
of $Nu$ as a function of $Pr$ for fixed $Ra=10^6$.
A power law fit to all available experimental data points \cite{cio97,hor98}
included in figure \ref{fig_nu_vs_pr} gives 
$Nu=(7.8\pm 0.5) Pr^{0.13\pm 0.02}$ which is in  agreement
with the theoretical expectation  
$Nu = 0.27 Ra^{1/4} Pr^{1/8} = 8.5 Pr^{1/8}$ in regime $I_l$. 
Leaving out the data points for water ($Pr=7$) which strictly
speaking already belongs to regime $I_u$ gives  a slightly larger
power law exponent and a slightly larger prefactor, $Nu=8.7 Pr^{0.16}$. 
Both exponent and prefactor are consistent 
with the theoretical expectation.

\subsection{Widths of the boundary layers}
\begin{figure}[htb]
\setlength{\unitlength}{1.0cm}
\begin{picture}(11,12)
\put(0.5,0.5)
{\psfig{figure=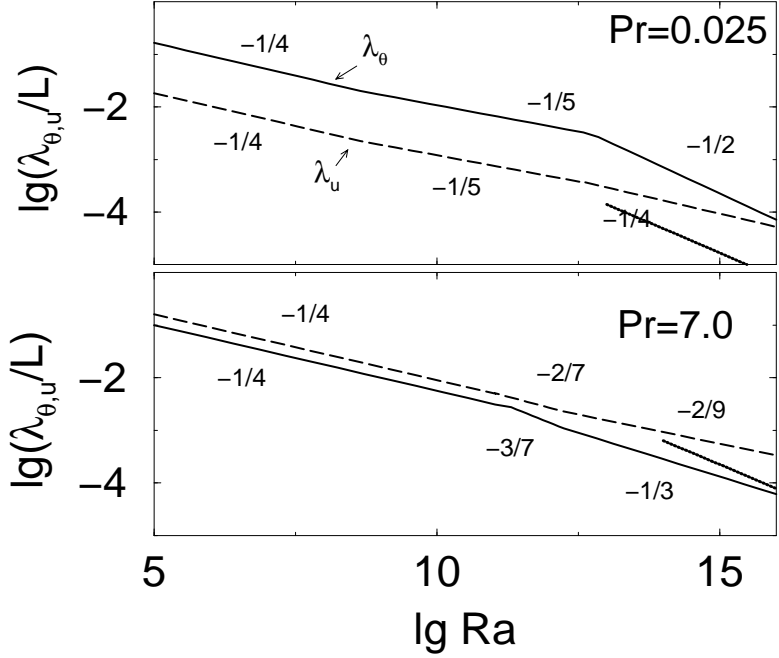,width=12cm,angle=-90}}
%{\psfig{figure=/usr/people/lohse/rb/bl.eps,width=12cm,angle=-90}}
\end{picture}
\caption[]{
Widths of the boundary layers $\lambda_\theta=0.5LNu^{-1}$ (solid)
and $\lambda_u =0.25 LRe^{-1/2}$ (dashed) for mercury ($Pr=0.025$, upper) and
water ($Pr=7.0$, lower). The  dotted lines show the thicknesses
$y_0$ of the viscous sublayers of a turbulent boundary layers,
calculated according to eq.\ (\ref{eqy0}). 
Such a width is expected beyond the 
nonnormal-nonlinear transition to turbulence of the laminar shear
BL (cf.\ section II-F).
}
\label{fig_bl}
\end{figure}

As stated above,
the present 
theory also gives the absolute widths of the thermal and the laminar viscous
boundary layers,
\begineq
\lambda_\theta = 0.5 L Nu^{-1},\label{newlamth}
\endeq 
\begineq
\lambda_u = 0.25 L Re^{-1/2}.
\label{newlamu}
\endeq
The results for the widths of the BLs for 
$Pr=0.025$ (mercury) and 
$Pr=7.0$ (water) are shown in figure \ref{fig_bl}. 
In both cases there are three regimes involved,
namely, $I_l$, $II_l$, and $IV_l$ for $Pr=0.025$
and $I_u$, $III_u$, and $IV_u$ for $Pr=7.0$. 
As expected, for the larger $Pr$ numbers the thermal
boundary layer is always nested in the viscous one
which agrees with the experimental observations \cite{bel94}.
For the lower $Pr$ numbers it is the other way round.

If the laminar BL becomes  turbulent, the Blasius estimate $\lambda_u
\sim L Re^{-1/2}$ for its width must be replaced by the thickness
of the turbulent BL. 
To give an idea about this length scale, we calculate the width $y_0$
of the viscous sublayer of the turbulent BL within the Prandtl theory
\cite{ll87}, applied to Couette flow. In the large $Re$ limit it holds
\cite{ll87,unpub}
\begineq
{y_0 \over L} = 1.38 {\log (k^2 Re)\over k^2 Re}. 
\label{eqy0}
\endeq
Here, $k=0.4$ is the experimentally known van K\'arm\'an constant. 
We have included $y_0/L$ in figure \ref{fig_bl} for the relevant large $Ra$. 
The turbulent viscous sublayer 
 is thiner  than the laminar BL.
One also notes that for the larger
Prandtl number $Pr=7$ (water) Shraiman and Siggia's assumption \cite{shr90}
of the thermal boundary layer being nested in the turbulent one is 
just fulfilled.
For lower $Pr$ this is not the case any more.

%caption1

Many experiments justify the identification  of the
thermal BL width with the inverse Nusselt number, $\lambda_\theta = 0.5 L Nu^{-1}$.
For a detailed discussion we refer to the review articles or to
Belmonte et al.\ \cite{bel94} or to the more recent work by 
Lui and Xia \cite{lui98}.

The situation is more complicated for the width of the kinetic BL 
$\lambda_u$. Its measurement is  experimentally difficult. 
Moreover, experimental results on $\lambda_u$ seem to 
exist only for regime I. 

Belmonte et al.\ \cite{bel94} tried to measure $\lambda_u$ 
in an indirect way, namely, through the detection of a spectral
cutoff frequency in gas convection which in water convection (at one
$Ra$) is found to have a similar dependence on the height $z$ in the 
RB cell as the velocity profile
$U(z)$. For $Pr\approx 1$ they found $\lambda_u \approx const$ in 
the regime $2\cdot 10^7 \le  Ra \le 2\cdot 10^9$ and 
$\lambda_u \sim L Ra^{-0.44 \pm 0.09}$ in 
$2\cdot 10^9 \le  Ra \le 10^{11}$. 
We have no idea about  the origin of the measured scaling exponents. 

More recently, Xin et al.\ \cite{xin96},
Xin and Xia \cite{xin97}, and Qiu and Xia \cite{qiu98}
measured the thickness of the kinetic BL 
 in a water cell in a more direct way. 
They define $\lambda_u$ as the distance from the wall at which the 
extrapolation of the linear part of the velocity profile $U(z)$
equals the maximum velocity $U= \hbox{max}_zU(z)$, the velocity of the large
scale wind. 
In the interval $2\cdot 10^8 \le  Ra \le 10^{10}$ they find
$\lambda_u \sim L Ra^{-0.16\pm 0.02}$ for the 
thickness of the top and bottom kinetic BL 
\cite{xin96,xin97}
and 
$\lambda_u \sim L Ra^{-0.26\pm 0.03}$ for the 
thickness of the  kinetic BLs at the side walls 
\cite{qiu98}.
The first exponent (for the top and the bottom plates)
is different from the value of this theory
$\lambda_u \sim L Ra^{-1/4}$.
We can only speculate on the origin of this  discrepancy. 
Perhaps, if $\lambda_u$ is defined as the distance of the velocity
maximum to the wall, the $Ra$-scaling would be different.
The power law for the thickness of the kinetic BLs at the
side walls, however, is  well in agreement with the expectation
$\lambda_u \sim L Ra^{-1/4}$.

In any case, the experimentally found very weak dependence of $\lambda_u$
on $Ra$ supports  the assumed {\it laminar}
 nature of the kinetic BL. If the width
of the  BL were 
identified with the width $y_0$ of the viscous sublayer of a {\it turbulent}
BL, one would expect a stronger dependence on $Re$, namely 
$y_0 \sim L \log(k^2 Re)/(k^2 Re)$ \cite{ll87,unpub}, 
i.e., when neglecting logarithmic corrections,
one would have 
$y_0 \sim L/Re \sim L Ra^{-1/2}$.

\subsection{Experimental evidence for the turbulence onset in the BL}
According to the presented theory with the chosen prefactors
the breakdown of laminarity in the shear BL happens at 
$Ra_{turb} \approx 10^{16}$ for $Pr=1$ and
$Ra_{turb} \approx 10^{14}$ for $Pr=0.025$.
These values are calculated from table \ref{tab_bounds}. 
Hitherto, there are no experiments for these regimes.

However, one may want to argue that the transition to a turbulent
shear BL may already occur earlier, be it because of a different
aspect ratio, or, in view of more recent work  \cite{eck99,bruno99b}, 
because the critical Reynolds number (which we
had assumed to be 420, cf. 
eq.\ (\ref{s2}))
is smaller, or because of a different
choice of the experimental input information to which the 
prefactors of the theory are
adopted. E.g., if one assumes a laminar shear
layer with width $\lambda_u = 1.72 L /\sqrt{Re}$ as suggested
by Landau and Lifshitz (ref.\ \cite{ll87}, section 39) for the
(related) case of a flat plate shear flow, the transition to
turbulence in the shear BL already occurs at $Ra_{turb} = 10^{13}$ 
(for $Pr=4$). We note that this is just at about that $Ra$ number 
where a marked transition in the (thermal) dissipative spectral power 
has been measured by a probe placed in the BL
\cite{pro91}. This transition was 
towards a weaker increase with $Ra$.

In the context of this section we also interpret the above mentioned 
recent experiment
by Ciliberto et al.\ \cite{cil99} in which  the boundary layers 
are disturbed by constructing a rough bottom plate with a mean roughness
comparable to the thermal boundary layer thickness. The experiments
are performed in water. According to the theory of this paper
one would expect larger bulk contributions to both the thermal and the kinetic
dissipation and therefore an earlier onset of regime 
$III^{\prime}_u$ and $IV^{\prime}_u$. 
Indeed, experimentally
the increase of $Nu$ with $Ra$ is found to be much steeper. For the experiment
described by Ciliberto et al.\ \cite{cil99} the data can be fitted to  power
laws 
$Nu \sim Ra^{0.35}$ or
$Nu \sim Ra^{0.45}$, 
depending on the features of the rough bottom and upper plate.

\section{Summary and conclusions}
We summarize the central ingredients of the theory
presented in this paper: 
The scaling laws for the Nusselt number and the 
Reynolds number are 
based on
the decomposition of the global thermal and  kinetic energy dissipation 
rates into their BL and bulk contributions.  These in turn 
are estimated from the 
dynamical equations, taking
the wind $U$ as the relevant velocity in the heat conduction cell. 
 The resulting estimates are inserted into  the rigorous  
relations eqs.\ (\ref{eq11}) and (\ref{eq12}) for the global 
kinetic and thermal energy 
dissipation, respectively. Four regimes arise, depending on whether the
bulk or the BL contributions dominate the two global dissipations. 
Each of the four regimes in principle
divides into two subregimes, depending on
whether the thermal BL (of width $\lambda_\theta$) 
or the kinetic BL (of width $\lambda_u$) is more extended.

In addition to these main regimes there is a range for very large $Pr$
numbers 
in which  the wind Reynolds number is  $\le 50$; here the whole
flow is viscosity dominated, and the theory looses its applicability.
There also is the range of very small $Pr$ numbers in which the $Nu$
goes down to
$Nu = 1$, and again the theory no longer holds. Finally, for large $Ra$ 
the laminar kinetic BL becomes turbulent.
Beyond turbulence onset we feel the flow is bulk dominated.

All scaling exponents follow from this theory. 
If one in addition introduces only four pieces of experimental information, 
also all the  prefactors can be  determined.
Therefore the theory
has predictive power not only for the power law exponents but 
also for the prefactors.
These, however, depend  on the chosen experimental 
information input.
To nail the prefactors 
more input 
information for various aspect ratios is necessary.

The  phase diagram of the theory, the main result of this work, is shown
in figure \ref{fig_sketch}. The power laws with the prefactors
based on the chosen experimental information 
are summarized in table \ref{tab_scal_laws}, the power laws of the
boundaries between the different regimes in table \ref{tab_bounds}.

A detailed comparison of the theoretical power law exponents
{\it and} the 
 prefactors with the experimental data gives reasonable 
 and encouraging agreement.
 We emphasize that to accurately account for the dependences
of $Nu$ and $Re$ on $Ra$ and $Pr$ 
single power laws are often not sufficient, as additive corrections
from neighboring regimes can be considerable.
This can be viewed as one of the main insights obtained in this paper. A particularly
striking example is that 
$Nu = 0.27 Ra^{1/4} + 0.038 Ra^{1/3}$ mimicks a $2/7$ power law scaling
over at least {\it nine} order of magnitude in $Ra$, see figure 
\ref{nu_vs_ra_log_2733}.

The theory also offers a possible explanation why a transition
to a steeper increase of $Nu$ vs $Ra$ is seen in the Chavanne et 
al.\ data \cite{cha97}, but not in the Chicago group data 
\cite{cas89,wu91a,pro91}.
It may be that in the large $Ra$  Chicago experiments the $Pr$ 
number was smaller
than in the Chavanne et al.\ \cite{cha97} measurements. Then 
for the Chicago  data
one had to expect a 
transition towards regime $IV_u$ where the $Ra$ scaling exponent
is $1/3$ and thus, as demonstrated, in superposition with the leading
$1/4$ exponent, is indistinguishable from a $2/7$
scaling. In the Chavanne et al.\ \cite{cha97} experiments, 
on the other hand, one
has the transition to the large $Pr$ number regime $III_u$ where
the $Ra$ scaling exponent is $3/7$ which is much better distinguishable
from $1/4$. Whether this explanation is true remains to be seen.

Finally, we want to stress and discuss one of the basic assumptions
of the theory, namely, that a large scale ``wind of turbulence'' exists,
defining the Reynolds number $Re = UL/\nu$, creating a shear BL
and stirring the turbulence in the bulk. 
Clearly, this assumptions breaks down in the shaded area 
in figure \ref{fig_sketch} beyond
the line $Re=50$ where the flow is viscosity dominated.
But even below this line we do not exclude that the convection rolls
break down and that the heat is exclusively transported by the 
fluctuations. E.g., for a water cell ($Pr\approx 7$) 
Tanaka and Miyata \cite{tan80} do not note the wind of turbulence,
in contrast to Zocchi et al.\ \cite{zoc90}, who do observe  the wind of
turbulence in their experiment with only a slightly lower
aspect ratio. Also all latest experiments 
with various aspect ratios 
doubtlessly detect the wind of turbulence 
\cite{bel94,xin96,qiu98,lui98} whose existence we therefore consider as
a weak -- and in particular controllable -- assumption. 

The present theory does {\it not} make any statement 
{\it how} the heat is transported from the bottom to the top, i.e.,
whether it is mainly convective transport or mainly transport 
through plumes \cite{bel94}. Both processes may contribute,
as both create thermal and viscous dissipation. 

For even larger Prandtl numbers 
$Pr \gg 7$,  the spontaneous formation of a wind of turbulence
may seem more and more unlikely. To initiate such a wind so that the
presented theory can be applied and results can be compared we suggest
to 
slightly tilt the RB cell,
thus breaking the symmetry and creating a preferred direction
for the wind of turbulence.  
If this wind of turbulence can be created, we are confident that 
the suggested theory holds. 
Presently, we do not have any experimental information which 
contradicts the theory. However, for further verification more
experiments will be valuable.

Clearly, this paper will not bring the turbulent RB 
heat conduction problem to an end.
But we hope that the presented new approach will restimulate
the discussion.

\vspace{1.5cm}

\noindent
{\bf Acknowledgments:}
D.\ L.\ gratefully acknowledges 
S.\ Ciliberto's hospitality during his visit in Lyon.
It was him 
who renewed our interest in Rayleigh-Benard
convection. We thank him and S.\ Cioni and X.\ Chavanne 
for various discussions and also 
for supplying us with their experimental data and allowing
us to reproduce them. We also thank G.\ de Bruin, L.\ Kadanoff,
V.\ Steinberg, 
 and A.\ Tilgner 
for hints and discussions. -- 
Support for this work by
the Deutsche Forschungsgemeinschaft (DFG) under the grant
Lo 556/3-1, by
the German-Israeli Foundation (GIF),  
and by the Chicago MRSEC-grant
is also acknowledged.
This work is  part of the research  programme of
the ``Stichting voor Fundamenteel Onderzoek der Materie (FOM)'', which
is    financially supported   by  the  ``Nederlandse  Organisatie voor
Wetenschappelijk Onderzoek (NWO)''.
\vspace{0.5cm}

\noindent
 e-mail addresses:\\
grossmann@physik.uni-marburg.de\\ 
lohse@tn.utwente.nl\\

%\vspace{1cm}

%\newpage

%\bibliography{literatur}

\end{document}